\documentclass[twocolumn]{aastex631}

\usepackage{newtxtext,newtxmath}
\usepackage{ae,aecompl}

\usepackage{graphicx}	
\usepackage{amsmath}	

\usepackage[dvipsnames]{xcolor}

\newcommand\msun{{\,M_\odot}}

\newcommand\zsun{{\rm \,Z_\odot}}

\newcommand{\cMpc}{~\mbox{comoving}~\mbox{Mpc}}

\newcommand{\cmci}{~\mbox{cm}^{-3}}

\newcommand{\K}{~\mbox{K}}

\newcommand{\traphic}{{\sc traphic}}

\newcommand{\gadgetthree}{{\sc p-gadget3}}

\newcommand{\Msun}{~\mbox{M}_{\odot}}

\newcommand{\Zsun}{~Z_{\odot}}

\newcommand{\kmsMpc}{~\mbox{km~s}^{-1}~\mbox{Mpc}^{-1}}

\newcommand{\music}{{\sc music}}
\newcommand{\cloudy}{{\sc cloudy}}
\newcommand{\rockstar}{{\sc rockstar}}

\newcommand{\hyperion}{{\sc hyperion}}
\newcommand{\yggdrasil}{{\sc yggdrasil}}

\newcommand{\nspace}{{ }}

\begin{document}

\title{How Massive Can a Population~III Starburst Be? Simulating the First Galaxies with High Lyman-Werner Background}

\author[0009-0000-8108-6456]{Tae Bong Jeong}
\affiliation{Department of Astronomy, The University of Texas at Austin, Austin, TX, 78712, USA}
\affiliation{Cosmic Frontier Center, The University of Texas at Austin, Austin, TX, 78712, USA}
\affiliation{Department of Astronomy \& Space Science, Kyung Hee University, 1732 Deogyeong-daero, Giheung-gu, Yongin-si, Gyeonggi-do 17104, Republic of Korea}
\correspondingauthor{Tae Bong Jeong}
\email{taebong.jeong@utexas.edu}

\author[0000-0003-2237-0777]{Alessandra Venditti}
\altaffiliation{Cosmic Frontier Center Prize Fellow}
\affiliation{Department of Astronomy, The University of Texas at Austin, Austin, TX, 78712, USA}
\affiliation{Cosmic Frontier Center, The University of Texas at Austin, Austin, TX, 78712, USA}

\author[0000-0003-0212-2979]{Volker Bromm}
\affiliation{Department of Astronomy, The University of Texas at Austin, Austin, TX, 78712, USA}
\affiliation{Cosmic Frontier Center, The University of Texas at Austin, Austin, TX, 78712, USA}
\affiliation{Weinberg Institute for Theoretical Physics, Texas Center for Cosmology and Astroparticle Physics, University of Texas at Austin, Austin, TX 78712, USA}

\author[0000-0001-6529-9777]{Myoungwon Jeon}
\affiliation{Department of Astronomy \& Space Science, Kyung Hee University, 1732 Deogyeong-daero, Giheung-gu, Yongin-si, Gyeonggi-do 17104, Republic of Korea}
\affiliation{School of Space Research, Kyung Hee University, 1732 Deogyeong-daero, Giheung-gu, Yongin-si, Gyeonggi-do 17104, Republic of Korea}

\author[0000-0003-4512-8705]{Tiger Yu-Yang Hsiao}
\affiliation{Department of Astronomy, The University of Texas at Austin, Austin, TX, 78712, USA}
\affiliation{Cosmic Frontier Center, The University of Texas at Austin, Austin, TX, 78712, USA}

\author[0000-0001-8519-1130]{Steven L. Finkelstein}
\affiliation{Department of Astronomy, The University of Texas at Austin, Austin, TX, 78712, USA}
\affiliation{Cosmic Frontier Center, The University of Texas at Austin, Austin, TX, 78712, USA}

\author[0000-0002-0302-2577]{John Chisholm}
\affiliation{Department of Astronomy, The University of Texas at Austin, Austin, TX, 78712, USA}
\affiliation{Cosmic Frontier Center, The University of Texas at Austin, Austin, TX, 78712, USA}


\begin{abstract}
Observing the first generation of Population~III (Pop~III) stars is one of the most demanding challenges in astronomy. Indeed, Pop~III stars are expected to predominantly form within faint minihalos at early times with a top-heavy initial mass function, resulting in efficient metal enrichment and a fast transition to Pop~II-dominated systems. However, recent surveys with JWST have identified galaxies at the end of the Epoch of Reionization (EoR) with possible signatures of significant Pop~III star formation even at these later times. We here explore the physical conditions required to produce massive Pop~III starbursts during the EoR, using cosmological radiation-hydrodynamic zoom-in simulations. We specifically focus on galaxies with a virial (dynamical) mass of $M_{\rm vir}\approx10^{8}\msun$ at $7\lesssim z\lesssim8$, i.e., the atomic-cooling halos that could be potential sites for such maximal Pop~III starbursts. In particular, we vary the strength of Lyman-Werner (LW) background radiation up to $J_{\rm LW}\leq10^4J_{21}$, further imposing a high star formation efficiency ($\epsilon_{\rm ff}=1.0$). Our results show that Pop~III starbursts, observable in strongly-lensed survey fields like GLIMPSE, can occur in the presence of a sufficiently high LW flux (with $\gtrsim10^3J_{21}$), leading to delayed, but intense Pop~III star formation. However, even for such high LW fluxes, the Pop~III starburst mass is limited to $M_{\star,\rm Pop~III}<10^6\msun$, as strong internal metal enrichment occurs after the first Pop~III supernova explosions within the simulated galaxies. While the conditions favoring observable Pop~III starbursts are expected to be rare, we anticipate that future and ongoing large-volume surveys leveraging gravitational lensing will detect multiple cases of Pop~III starbursts in the EoR.
\end{abstract}

\keywords{Early universe (435), Galaxy formation (595), High-redshift galaxies (734), Hydrodynamical simulations (767), Population III stars (1285), James Webb Space Telescope (2291)}

\section{Introduction}

\par
Detecting the direct signature of the first generation of stars, the so-called Population~III (Pop~III), is one of the key objectives in modern astronomy (e.g., \citealp{Barkana2001, Bromm2009, Bromm2013, Klessen2023}). Although most Pop~III stars likely formed and ended their lives by $z \gtrsim 15$, theoretical studies suggest that a subdominant Pop~III component could survive until the end of the Epoch of Reionization (EoR) at $z \gtrsim 5$ (e.g., \citealp{Tornatore2007, Johnson2010, Maio2010, Xu2016, Liu2020MNRAS, Venditti2023, Venditti2024, Venditti2024_PISN, Venditti2025, Ventura2024MNRAS, van_Veenen2025, Zier2025}). 
Due to the patchy nature of metal enrichment, pristine or extremely low-metallicity gas clouds might survive in rare pockets that are not reached by metal-enriched outflows from nearby structures \citep[e.g.,][]{Jeon_2017,Jaacks2019, Brauer2025}.
Still, the direct detection of Pop~III stars or galaxies poses a demanding challenge. As Pop~III stars are thought to exhibit a top-heavy initial mass function (IMF), they are predicted to be short-lived, and also typically form within unpolluted minihalos in small groups of stars (e.g., \citealp{Greif2011, Hirano2017, Prole2022, Klessen2023, Lee2024, Jeon2026}). Therefore, observing the faint direct signatures of Pop~III stars at $z\gtrsim 20$, when the cosmic dark ages end, is beyond the capabilities of even the James Webb Space Telescope (JWST; see \citealt{Schauer_ULT2020,Zackrisson2024}). The alternative approach of directly detecting Pop~III stars or galaxies at somewhat later times, when metal-enriched star formation dominates, on the other hand, is akin to ``finding a needle in a haystack' (\citealp{Venditti2023}).
\par
However, using frontier facilities such as the JWST, possibly involving further flux amplification through gravitational lensing, a few extremely metal-poor galaxy candidates in the EoR (e.g., \citealp{Cullen2023B, Vanzella2023, Chemerynska2024, Maiolino2024b, Wang2024, Fujimoto2025, Hsiao2025, Nakajima2025, Morishita2025, Willott2025, Cullen2025, Maiolino2026, Ubler2026}), and one candidate in the post-EoR era (\citealp{Cai2025, Mondal2025}), have been detected. Given the presence of significant nebular continuum and He~II~1640\,\AA\,flux, characteristic consequences of hard ionizing spectra from massive Pop~III stars (e.g., \citealp{Bromm2001, Schaerer2002,Schaerer2003, Trussler2023,Katz2023,Venditti2024}), these candidates could be the first instances of direct Pop~III signatures.
For example, utilizing an extreme gravitational lensing boost ($\mu \gtrsim 98$),  \citet{Vanzella2023} reported LAP1-A, observed at $z \sim 6.6$ behind a lensing cluster. Moreover, \citet{Nakajima2025} estimated its stellar mass as $M_{\star} <  2700 \msun$ for LAP1-B (which is inferred to reside in the same overall halo as LAP1-A), which could represent the first episode of star formation and chemical enrichment for the possible high-$z$ progenitor of ultra-faint dwarf (UFD) galaxies in the local Universe \citep[e.g.,][]{Frebel2012,Jeon_2017}. This object shows He~II~1640\,\AA\,emission \citep[e.g.,][]{Bromm2001,Schaerer2002}, as well as low metal-line fluxes (CIV 1549\,\AA \, and [O~III] 5007\,\AA), compared with the intensity of H lines. However, the continuum of this source remains undetected, rendering its physical properties uncertain.
\par
Furthermore, \citet{Morishita2025} discovered AMORE6, a Pop~III galaxy candidate at $z= 5.725$, with stellar mass of $M_{\star} \approx 5.6\times 10^5 \msun$. Although follow-up observations are needed for confirmation, this candidate exhibits a high H$\beta$ line flux, without detectable [O~III] 4959, 5007\AA, resulting in a high probability for a Pop~III dominated galaxy.
Lastly, \citet{Cai2025} reported a Pop~III galaxy candidate at $z \approx 3.19$ with stellar mass $M_{\star} = 6.1 \times 10^5 \msun $, dominated by very young ($t_{\rm age} \approx 2$ Myr) stars. Although they could not detect the He~II 1640\,\AA \, line due to the presence of strong sky emission lines, they did detect strong Ly$\alpha$ and Balmer features without a contribution from metal lines. This indicates low metallicity and the presence of a hard, ionizing continuum, which could be the direct signature of Pop~III stars \citep[e.g.,][]{Johnson_2009}.
\par
Since Pop~III galaxy candidates during the EoR are detected with a range of stellar masses, significant questions arise regarding their formation and evolution, such as how Pop~III galaxies could have avoided metal enrichment earlier on, and how massive they can be. The most crucial condition for a Pop~III starburst is the survival of primordial cold dense gas in their host halos. Given that Pop~III star formation requires extremely low metallicity, even a single supernova (SN), or a few such events, inside the first minihalos could establish the conditions where cooling processes become efficient,  driving a rapid transition to Population~II (Pop~II) star formation (e.g., \citealp{Jeon_2014}).
\par
Previous theoretical studies have suggested that Pop~III star formation could be possible within more massive halos, exceeding the virial mass of the canonical molecular-cooling regime (e.g., \citealp{Fisk2025,Hegde2025,Riaz2022,Venditti2025,Storck2025}). Additionally, \citet{Venditti2025} suggested that highly bursty Pop III star formation in halos up to the atomic-cooling regime ($M_{\rm vir} \sim 10^8 \msun$ at $z \sim 5.6-6.6$), or efficient Pop III formation within heavier host halos ($M_{\rm vir} \lesssim 10^{10} \msun$), would be necessary to explain the bright Pop~III galaxy candidates recently discovered in the EoR (\citealp{Fujimoto2025, Fujimoto2025b, Morishita2025}).
\par
The suppression of star formation in molecular-cooling halos could be brought about by a strong Lyman-Werner (LW) flux from nearby sources or an external background (e.g., \citealp{Agarwal2012, Regan2014, Visbal2017, Wolcott-Green2017, Schauer2021, Prole2024, Fisk2025, van_Veenen2025}), which dissociates molecular hydrogen \citep{Haiman1997}. As a consequence, cooling inside minihalos is inhibited, until their virial temperature exceeds $T_{\rm vir}\approx8000\rm\, K$, where atomic hydrogen can provide a strong cooling channel. For example, \citet{Visbal2017} investigated the formation of Pop~III galaxies in the presence of isotropic photoionization and LW radiation. Although star formation was not explicitly modeled, they inferred Pop~III stellar masses inside their simulation box up to $\sim 10^6\msun$. Similarly, \citet{Schauer2021} suggested that even low values of the LW background\footnote{We here employ the usual LW flux calibration in terms of the standard value of $J_{21}=10^{-21}\,{\rm erg\,s^{-1}\,cm^{-2}\,Hz^{-1}\,sr^{-1}}$.} from nearby sources ($J_{\rm LW} \leq 0.1 J_{21}$) could delay the first star formation episodes, causing them to occur in halos with slightly larger virial mass compared to the typical mass of a molecular cooling halo, in the range of $10^6 \msun\lesssim M_{\rm vir} \lesssim 10^7\msun$.
\par
In this study, we further investigate the formation and evolution of Pop~III galaxies during the EoR. In particular, we ask: ``How massive can a Pop~III starburst be?'' To achieve this goal, we perform a series of cosmological radiation hydrodynamic zoom-in simulations, targeting halos with virial mass $M_{\rm vir} \approx 10^8 \msun$ at $7 \lesssim z \lesssim 8$ immersed in a LW background of various intensities. 
Based on our simulation results, we derive synthetic observables to assess the detectability of Pop~III systems in recent JWST surveys. We will focus on the combination of delayed star formation due to LW radiation and its bursty nature as key physical concepts to understand the formation of Pop~III galaxies in the EoR, probing their maximum stellar mass.
\par
This paper is organized as follows. In Section~2, we describe the numerical methodology, followed by a presentation of the detailed simulation results in Section~3. In Section~4, we discuss the observable properties derived from our simulations, along with the caveats and limitations of our work. Finally, we provide a summary and conclusions in Section~5. For consistency, all distances are given in physical (proper) units, unless noted otherwise.

\section{Numerical methodology}
\label{Sec:Method}
This study follows the numerical methodology detailed in \citet{Jeong2025}. Here, we briefly summarize our simulation and post-processing methodologies, focusing on aspects of the subgrid physics that are of particular relevance for the current study, specifically the prescriptions for star formation and LW radiation feedback. 
\par

\subsection{Simulations and post-processing}
We have performed radiation hydrodynamic zoom-in simulations using a modified version of the N-body/TreePM Smoothed Particle Hydrodynamics (SPH) code \gadgetthree \nspace (\citealp{2001NewA....6...79S}, \citealp{Springel_2005}). For the cosmological parameters, we adopt the matter density parameter $\Omega_{\rm m} = 1 - \Omega_{\rm \Lambda}= 0.265$, baryon density $\Omega_{\rm b}=0.045$, Hubble constant $H_{\rm 0} = 71 \kmsMpc$, and normalization parameter $\sigma_{\rm 8} = 0.81$  (\citealp{Planck2016}). The multi-scale initial conditions are generated using the cosmological initial condition code \music \nspace (\citealp{Hahn_2011}). We initially perform dark matter-only simulations at a lower resolution using $128^3$ particles within a box of $L_{\rm box}=6.25 h^{-1} \cMpc$, and identify our target halo using the halo-finder code \rockstar \nspace (\citealp{Behroozi_2013}). After selecting the target halos, we conduct successive refinements on the particles within 5$R_{\rm vir}$ at the end of the simulations, where $R_{\rm vir}$ denotes the virial radius of the halos. This process results in refined dark matter (DM) and SPH particle masses of $m_{\rm DM} \approx 2500\Msun$ and $m_{\rm SPH} \approx 500\Msun$, respectively, corresponding to an effective resolution of $2048^3$.
\par
Our simulations start at $z = 125$ and terminate within the range $6.5 \lesssim z \lesssim 7.0$, to compare our results with Pop~III galaxy candidates at the EoR. The adopted softening length in our simulations is $\epsilon_{\rm soft} \sim 30$ physical pc, which is kept constant across all redshifts for DM and stellar particles. Additionally, as in \citet{Jeong2025}, we use adaptive softening lengths for the gas particles, chosen to be proportional to the SPH kernel length, with a minimum value of $\epsilon_{\rm gas, min} = 2.8 \rm \, pc$. At each timestep, we solve the non-equilibrium rate equations for the primordial chemistry network involving thirteen atomic and molecular species, H, H$^+$, H$^{-}$, H$_2$, H$^{+}_{2}$, He, He$^+$, He$^{++}$, e$^{-}$, HD, HD$^{+}$, D, and D$^+$. In addition to primordial cooling, our simulations incorporate metal-line cooling processes for carbon, oxygen, silicon, magnesium, neon, nitrogen, and iron. The cooling rates for these elements are determined using the photoionization package \cloudy \nspace (\citealp{1998PASP..110..761F}). 
\par
After the formation of stars in the simulations, the photoionization feedback is followed using the radiative transfer (RT) module \traphic \nspace (\citealp{Pawlik_2011}), and we enabled photoionization feedback from both Pop~III and Pop~II stars.
SN feedback is triggered after a $t_{\rm age}=3\rm \,Myr$ life cycle of star particles, implemented as a thermal energy injection mechanism. We adopt the method proposed by \citet{DallaVecchia2012} to avoid the over-cooling problem, limiting the number of gas particles receiving the SN energy to $N_{\rm ngb} = 1$. We also consider the chemical enrichment from both Pop~III and Pop~II stars, using the method described in \citet{Wiersma2009}. For Pop~III stars, we adopt nucleosynthetic metal yields and remnant masses from \citet{Heger2010} for core-collapse supernovae (CCSNe) and from \citet{Heger2002} for pair-instability supernovae (PISNe). For Pop~II stars, we calculated the elemental yields using the metallicity-dependent table for CCSNe from \citet{Portinari1998}. We also include yield contributions from asymptotic giant branch (AGB) stellar winds \citep{Marigo2001}) and from Type~Ia SNe (\citealp{Barris2006}).
Metals ejected from dying stars are dispersed into neighboring gas particles ($N_{\rm ngb} = 48$), and then transported from the ISM to the IGM by solving the diffusion equation (see, e.g., \citealp{Greif2009}).
\par
Based on the simulation outputs, we derive synthetic observations through a post-processing pipeline that integrates {\sc fsps} \citep{Conroy2010}, \yggdrasil \, (\citealp{Zackrisson2011}), \hyperion \, (\citealp{Robitaille2011}), and \cloudy \, (\citealp{Ferland2017}). 
To model the IGM attenuation, we adopt the analytic model of \citet{Inoue2014}, which provides the neutral hydrogen optical depth as a function of redshift and wavelength for the Lyman-continuum ($\lambda \leq 912$\,\AA) and Lyman-series ($912$\,\AA $\leq \lambda \leq 1216 $\,\AA) radiation.

\subsection{Star formation}
\label{subsection:SF}
We consider the formation of Pop~III stars, which are the first generation of stars with extremely low metallicity, and subsequently of Pop~II stars that are formed from gas clouds enriched by Pop~III SN explosions. Stars are formed when gas particles exceed a threshold in hydrogen number density of $n_{\rm H, thr} = 100 \cmci$, while having temperatures of $T_{\rm thr} < 10^3 \rm \,K$. We implement a stochastic conversion of gas particles into star particles, where the rate follows the Schmidt law (\citealp{Schmidt1959}). In particular, gas particles are converted into stars according to $\dot{\rho}_{\star} = \rho / \tau_{\star}$, where $\rho$ represents the gas density, and $\tau_{\star}$ is the star formation timescale, defined as $\tau_{\star} = \tau_{\rm ff}/ \epsilon_{\rm ff}$. Here, $\epsilon_{\rm ff}$ denotes the star formation efficiency, and the free-fall time is given by $\tau_{\rm ff} = \left[3\pi/(32G\rho) \right]^{1/2}$. During a given simulation timestep, $\Delta t$, gas particles satisfying the star formation thresholds are converted into star particles only if a randomly generated number between 0 and 1 is less than $\rm min(\Delta t / \tau_{\star}, 1)$.

With these definitions, we can write the star formation timescale, as follows:
\begin{equation}
    \tau_{\star} = \frac{\tau_{\rm ff}(n_{\rm H, thr})}{\epsilon_{\rm ff}} = 201.2 \, \rm Myr \, \Big( \frac{\epsilon_{\rm ff}}{0.01} \Big)^{-1} \it \, \Big(\frac{n_{\rm H, thr}}{\rm 500 \cmci}\Big)^{\rm -1/2}.
    \label{eq:SF_timescale}
\end{equation}
In our default set-up, we use the same star formation efficiency for both Pop~III and Pop~II stars, $\epsilon_{\rm ff}=0.01$, which is a typical value observed in the local Universe (e.g., \citealp{Leroy2008}). When a gas particle satisfies these conditions, it is converted into a collisionless star particle with an initial mass of $m_{\star}=500\Msun$. Due to the limited numerical resolution, we treat our star particles as a single stellar population (SSP), with masses extracted from the assigned IMFs, rather than as individual stars.
\par
Pop~III stars form in (nearly) metal-free gas clouds, where the primary cooling agent is molecular hydrogen ($\rm H_{2}$). This process is relatively inefficient, allowing temperatures to drop only to about $T \approx 200$~K, increasing the probability of Pop~III stars being more massive ($\gtrsim 100\msun$) (e.g., \citealp{Bromm2013, Klessen2023}). Theoretical studies incorporating radiative feedback from protostars and disk fragmentation proposed the formation of multiple stellar systems, with less massive stars reaching up to several tens of solar masses (e.g., \citealp{Clark2011, Stacy2016, Sugimura2020, Latif2022}). In our simulations, Pop~III stars are formed while the metallicity of gas particles remains below a metallicity threshold\footnote{The precise value of this `critical metallicity' is uncertain, with possibly different thresholds for triggering low-mass star formation and the vigorous fragmentation required for cluster formation \citep[e.g.,][]{Safranek2014, Sharda2022, Smith2024}}. of $Z_{\rm thr} =10^{-5.5}\Zsun$ (e.g., \citealp{Omukai2000, Schneider2010, Safranek-Shrader2016, Sharda2022, Smith2024}). For Pop~III, we impose a top-heavy IMF, described by the functional form: $\phi_{\rm PopIII}(m) = dN/d\log m\approx m^{-\alpha}$, with a slope of $\alpha=1.0$ across the mass range $[m_{\rm min},m_{\rm max}]=[10\msun,150\msun]$.
\par
One way to achieve Pop III starbursts within the simulated galaxies is by simply increasing $\epsilon_{\rm ff}$. Specifically, we increase $\epsilon_{\rm ff}$ to 1.0, two orders of magnitude higher than the value used in our default settings, thus assuming efficient star formation inside our target halo. We expect that this increase in $\epsilon_{\rm ff}$ would result in bursty star formation (e.g., \citealp{Dekel2023, Jeong2025, Andalaman2025, Wang2025}), thereby allowing a rapid starburst, which concludes before the most massive stars in the target halo die (after $t_{\rm age} \approx 3\rm \,Myr$), in turn triggering vigorous SN feedback.
For instance, \citet{van_Veenen2025} inferred from their simulations that the cloud-scale star formation efficiency increases with the strength of the LW background. At the same time, they also suggested that the emerging stellar cluster exposed to a high LW background ($J_{\rm LW} = 5000\,J_{21}$) may act to seed massive BH formation. Given these uncertainties in high-LW flux scenarios, we emphasize that our results with $\epsilon_{\rm ff} = 1.0$ should be interpreted as an upper limit for a Pop~III starburst, representing the maximum achievable efficiency.
\par
At the end of their life, Pop III stars explode as either conventional CCSNe ($8\msun \leq M_{\star} \leq 40 \msun$) or PISNe ($140 \msun \leq M_{\star} \leq 260 \msun$), enriching the surrounding gas particles. If gas particles exceed $Z_{\rm thr}$ and satisfy the star formation thresholds, they are converted into Pop~II stars. Pop~II stellar particles are also formed with an initial mass of $M_{\star} = 500 \msun$, following a Chabrier IMF (\citealp{Chabrier2003}) within the mass range of $[m_{\rm min}, m_{\rm max}] = [0.1\msun, 100 \msun]$.

\subsection{Lyman-Werner Radiation}
\label{LW}
LW radiation from a combination of sources such as massive stars or black holes is capable to photo-dissociate $\rm H_{2}$ and inhibit its formation, thus suppressing the main cooling channel for pristine gas clouds within Pop~III host halos (e.g., \citealp{Johnson2008, Agarwal2012}). The mean intensity of the LW background is believed to follow the overall growth of cosmic structure, starting to build up at $z \approx 30$, reaching a peak of $J_{\rm LW}/J_{21} = 1 - 10$ at $z \approx 6 - 7$, and then steadily decreasing down to $z = 0$ (e.g., \citealp{Greif2006, Haardt2012, Johnson2013, Faucher-Giguere2020}).
\par
However, due to the patchy nature of both the cosmic reionization and metal enrichment histories, pristine pockets may persist in overdense regions immersed in an extreme LW radiation field, due to a high concentration of nearby sources (e.g., \citealp{Agarwal2012, van_Veenen2025}).
For example, \citet{Fisk2025} investigated the effect of radiation from active galactic nuclei (AGN) on the formation of Pop~III clusters using cosmological simulations. Adjusting the distance from the AGN to a given target halo, they suggest that star formation could be suppressed until a host halo has grown to $M_{\rm vir} \sim 10^9 \msun$, which could engender the formation of massive Pop~III clusters or of a direct-collapse black hole (DCBH). Also, \citet{Sullivan2025} investigated the formation of a supermassive star (SMS) in the presence of additional local LW sources, concluding that even with extreme values for the ``internal'' LW radiation, SMSs or massive BH seeds are unlikely to form. However, they also suggest the possibility that metal-free halos might be subject to extreme LW flux from nearby sources.
\par
For simplicity, in this study we model the local increase in LW radiation from multiple nearby sources as a uniform background expressed by $J_{\rm LW}(z)$ (applied to the central zoom-in region), adopting a simple step function form depending on redshift:

\begin{equation}
\label{LW-eq}
    J_{\rm LW} (z)/J_{21} = 
    \begin{cases}
    0 \qquad \qquad \qquad \qquad \; \, \rm for \, \it z > \rm 30,
    \\
    \bar{J}_{21, 0} \qquad \qquad \qquad \quad \rm for \, \it z \leq \rm 30,
    \end{cases}
\end{equation}
where $\bar{J}_{21, 0}$ is the mean specific intensity of the LW flux in units of $10^{-21}\rm erg\, s^{-1}\,cm^{-2}\, Hz^{-1}\, sr^{-1}$. In our simulation sets, we vary the strength of $\bar{J}_{21, 0}$ from 0 to $10^4$ to bracket all possibilities, including extreme cases for the photodissociation radiation from nearby sources \citep[e.g.,][]{Ahn2009}.

\par
Finally, we include dimensionless factors $f_{\rm shield,H_{2}}$, $f_{\rm shield,HD}$, and $f_{\rm shield,H_{2},HD}$, which describe the local flux attenuation from the self-shielding of $\rm H_2$ and HD, as well as from shielding of HD by $\rm H_2$. Specifically, we compute the self-shielding factors $f_{\rm shield,H2}$ and $f_{\rm shield,HD}$ using the equation from \citet{Wolcott-Green2011a} and \citet{Wolcott-Green2011b}:
\begin{equation}
    \begin{split}
        f_{\rm shield}(N_{i}, T) = \frac{0.965}{(1+x/b_5)^\alpha}+\frac{0.035}{(1+x)^{0.5}} \, \\
        \times \exp[-8.5 \times 10^{-4}(1+x)^{0.5}],
    \end{split}
\end{equation}
where $x \equiv N_i/5 \times10^{14} \rm \,cm ^{-2}$, and $N_i$ is the column density of the molecular species. In this equation, $b_5 \equiv b/10^5 \rm \,cm\,s^{-1}$, where $b \equiv \sqrt{2k_BT/m_p}$ is the Doppler broadening parameter, $m_p = 2m_{\rm H}$ the mass of the molecular species, and $\alpha = 1.1$.
We also evaluate $f_{\rm shield,H_{2},HD}$ as:
\begin{equation}
    f_{\rm shield,H_{2},HD} = \frac{1}{(1+x)^{0.238}}\,\exp{(-5.2\times10^3 x)}, 
\end{equation}
where $x \equiv N_{\rm H_{2}}/2.34 \times 10^{19} \rm \,cm ^{-2}$. We estimate the column density $N_i$ as $N_i= n_i L_{\rm char}$, where $n$ is the number density and $L_{\rm char}$ is the characteristic length scale, set to the local Jeans length (an approximation appropriate for self-gravitating systems, that performs well in comparison with more sophisticated approaches, e.g., \citealp{Wolcott-Green2011a}).

\par
Note that our simulations only run down to $z = 7.0$, when the Universe is not yet fully ionized. Therefore, we have not adopted an ionizing (hard) UV background in our default simulations (but see Section~\ref{subsection:Reionization}), and only include the effect of a soft UV (LW) background (e.g., \citealp{Becker2021, Becker2024, Lewis2022}).

\subsection{Summary of simulation suite}

\begin{table}[]
    \centering
    \begin{tabular}{c|c|c|c|c}
         \hline
         \textbf{Set name} & \textbf{$\bar{J}_{21,0}$} & \textbf{$\epsilon_{\rm ff}$} & $M_{\rm vir}$ at first SF & \textbf{Group name} \cr
         \hline \hline
         {\sc LW0E001} & 0 & 0.01 & $4.77 \times 10^6 \msun$ & NoLW \cr
         \hline
         {\sc LW0E100} & 0 & 1.0 & $4.07 \times 10^6 \msun$ & NoLW \cr
         \hline
         {\sc LW1E100} & 1 & 1.0 & $4.74 \times 10^7 \msun$ & Low \cr
         \hline
         {\sc LW10E100} & 10 & 1.0 & $7.55 \times 10^7 \msun$ & Low \cr
         \hline
         {\sc LW100E100} & 100 & 1.0 & $7.96 \times 10^7 \msun$ & Intermediate \cr
         \hline
         {\sc LW1e3E100} & 1,000 & 1.0 & $9.23 \times 10^7 \msun$ & High \cr
         \hline
         {\sc LW1e4E100} & 10,000 & 1.0 & $1.04 \times 10^8 \msun$ & High \cr
         \hline
    \end{tabular}
    \caption{Summary of simulation parameters. Column (1): Name of run. Column (2): Strength of LW background in units of $10^{-21}\rm \,erg\,s^{-1}\,cm^{-2}\, Hz^{-1}\, sr^{-1}$.  Column (3): Star formation efficiency $\epsilon_{\rm ff}$. Column (4): Halo virial mass at onset of star formation. Column (5): Assigned group based on the strength of $\bar{J}_{21, 0}$.}
    \label{tab:simulation_summary}
\end{table}

To investigate the theoretical upper limit for the mass formed in a Pop~III starburst, we carry out seven simulation runs, targeting halos with $M_{\rm vir} \approx 10^8 \msun$ at $z = 7$, varying two main parameters of our sub-grid physics recipes, the values of $\bar{J}_{21, 0}$ and $\epsilon_{\rm ff}$.
The key features of our simulation runs are briefly summarized in Table~\ref{tab:simulation_summary}, including the virial mass of each halo at its first star-formation episode.
We separated our simulation sets into four different groups, based on the strength of $\bar{J}_{21, 0}$: \textbf{NoLW} (LW0E001, LW0E100), \textbf{Low} (LW1E100, LW10E100), \textbf{Intermediate} (LW100E100), and \textbf{High} (LW1e3E100, LW1e4E100).

\section{Simulation Results}
\label{Result:Sim}
In this section, we present our results for the formation and evolution of Pop III-hosting galaxies, with particular emphasis on the impact of the assumed LW flux arising from multiple local sources. Specifically, we examine the mass assembly of the simulated galaxies (Section~\ref{Sim:Mass}), the physical properties of massive Pop~III starbursts (Section~\ref{Sim:Gas}), and the evolution of their gas-phase metallicity (Section~\ref{Sim:Met}). 

\subsection{Assembly histories}
\label{Sim:Mass}

\begin{figure*}
    \centering
    \includegraphics[width = 170mm]{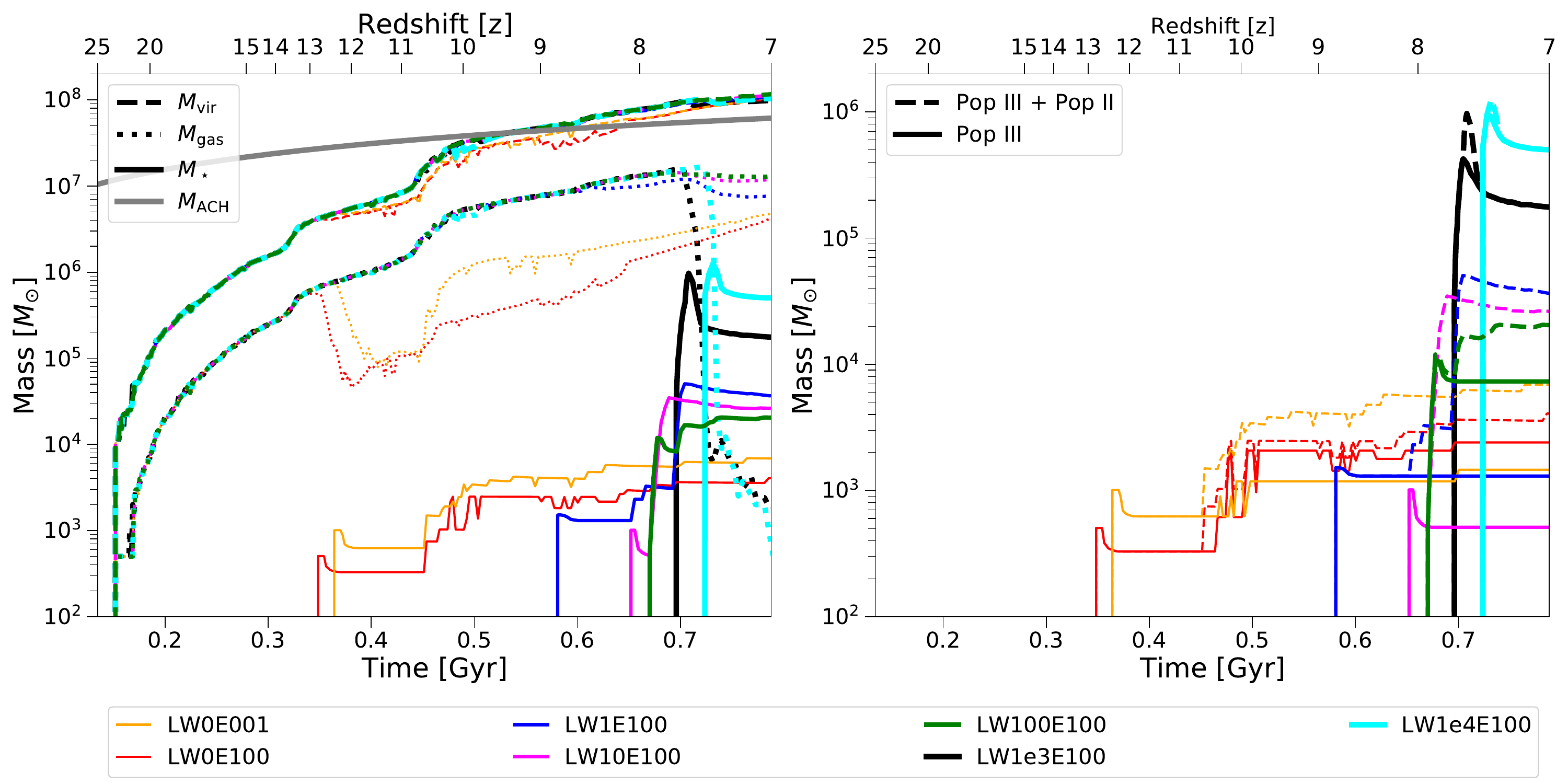}
    \caption{Mass evolution within our simulated galaxies. The left panel showcases the evolution of virial mass ($M_\mathrm{vir}$, dashed), gas mass ($M_\mathrm{gas}$, dotted), and stellar mass ($M_\star$, solid) within the virial radius of the target halo as a function of time (or redshift), with the thick gray line indicating the threshold virial mass for an atomic-cooling halo ($M_\mathrm{ACH}$; \citealt{Prole2024}) as a reference. The right panel further shows the evolution of the stellar mass within the target halo, separating the total stellar mass (dashed) from its Pop~III contribution (solid). Different simulation runs are shown in different colors, with increasing line widths indicating increasing values of the strength of the LW flux ($\bar{J}_{21,0}$) in different simulation groups (\textbf{NoLW}, \textbf{Low}, \textbf{Intermediate}, \textbf{High}). 
    }
    \label{fig:mass_evolution}
\end{figure*}

\par
Fig. \ref{fig:mass_evolution} presents the composite set of mass evolution histories for our simulated galaxies, including the virial mass ($M_{\rm vir}$), gas mass ($M_{\rm gas}$) and stellar mass ($M_{\star}$) enclosed within the virial radius of each halo, as well as the Pop~III contribution to the total stellar mass.
Since we adopt the same initial conditions for the entire simulation suite, the virial mass follows a similar trend regardless of the adopted baryonic physics setup, reaching $M_{\rm vir} \sim 10^8 \msun$ by $z=7.0$ in all simulations. On the other hand, the assembly of baryonic matter -- gas and stars -- differs greatly in each run, primarily due to the diverse environmental conditions linked to different LW flux intensities.
\par
Within the \textbf{NoLW} group, the LW0E001 run forms the first Pop~III star cluster at $z \approx 12.4$ when the virial mass of the simulated galaxy has reached $M_{\rm vir} \approx 4.77 \times 10^6 \msun$, which is in the typical mass range of molecular-cooling halos (e.g., \citealp{Bromm2009, Jeon_2019, Jeong2025}). The increase in $\epsilon_{\rm ff}$ in the LW0E100 run has a minor impact on baryonic mass assembly, only resulting in a slight anticipation of the first star-formation episode to $z \approx 12.79$ and a slight increase in the total stellar mass produced, due to the shorter star-formation timescale in Equation~\ref{eq:SF_timescale}. In both runs, star formation is quenched soon after by stellar photoionization feedback, and SN explosions also occur after a short time delay ($t_\mathrm{age} = 3 \rm \,Myr$, see Section~\ref{subsection:SF}), triggering an outflow that ejects $\sim 90\%$ of the gas mass.
Gas is only reaccreted in the halo after a prolonged quenching time ($\Delta t_{\rm delay} > 100 \rm \,Myr$).
As a result, the total stellar mass within the simulated galaxies at the end of the \textbf{NoLW} runs remains low ($M_{\star} \lesssim 10^4\msun$.)
\par
The simulated galaxies in the \textbf{Low} group show markedly different evolution histories. The LW flux dissociates $\rm H_{2}$ molecules, delaying star formation until the virial mass exceeds the atomic-cooling limit, $M_{\rm vir} \approx 4.7 \times 10^7 \msun$ ($M_{\rm vir} \approx 7.6 \times 10^7 \msun$) for the LW1E100 (LW10E100) run. Afterwards, the central gas clouds can finally collapse due to efficient atomic cooling, until they become dense enough to self-shield against radiation and form H$_2$; molecular cooling then allows them to reach $T_{\rm gas} \lesssim10^3\rm \,K$. As a result, the first star-formation episode occurs at $z \approx 8.8$ and $z \approx 8.1$ for the LW1E100 and LW10E100 runs respectively, with a delay time with respect to the \textbf{NoLW} group of $\Delta t_{\rm LW} \sim 200-300 \rm \, Myr$.
SN feedback from the first generation of stars disrupts the cold dense gas at the center of the galaxies, halting star formation. 
However, due to the deeper potential well of the halos at the time of the first SN explosions compared with the \textbf{NoLW} group, they can retain most of their gas mass. As a consequence, the second generation of Pop~II stars forms after a shorter time delay of $\Delta t_{\rm delay} \sim 75$~Myr (22 Myr) for the LW1E100 (LW10E100) run.
Most of the stellar mass is formed in this second, short burst of $\Delta t_\mathrm{burst} \sim \rm 15\, Myr$, reaching $M_{\star} \lesssim 5\times 10^4 \msun$, after which star formation is quenched by SN feedback until the end of the simulations.
\par
\begin{figure*}
    \centering
    \includegraphics[width = 170mm]{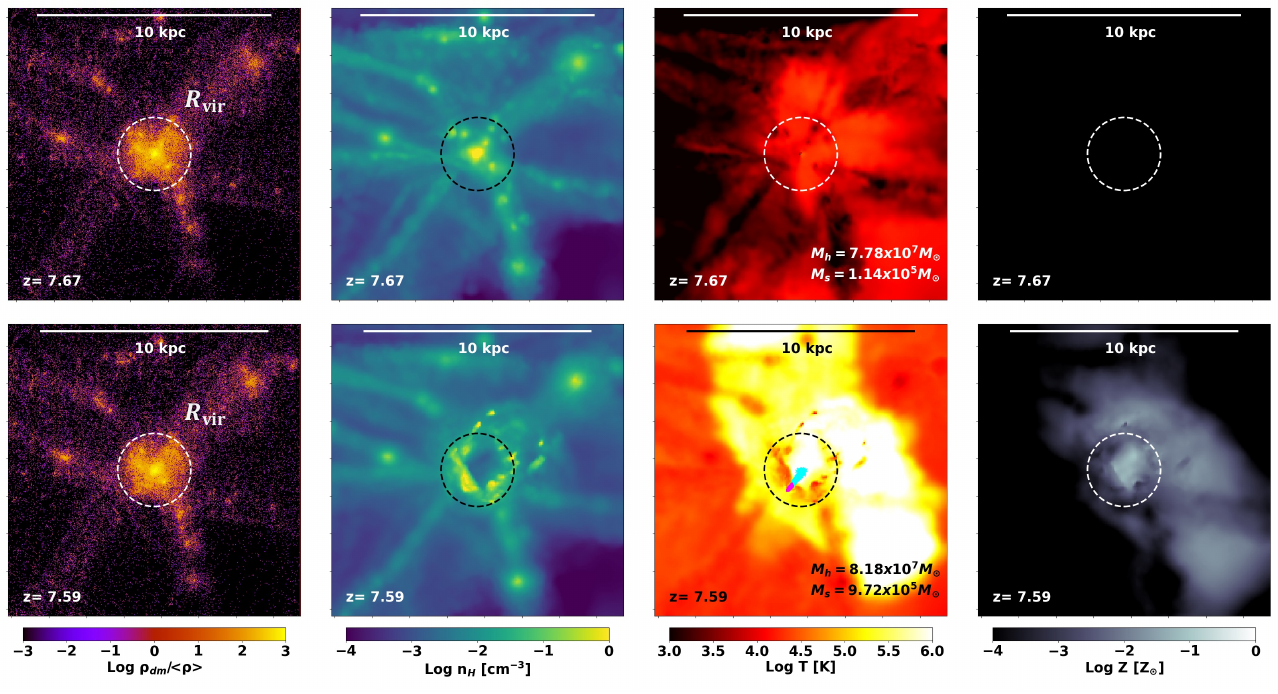}
    \caption{Spatially resolved properties of the simulated galaxy from the LW1e3E100 run at $z \approx 7.67$ (top row), when the first star formation occurred, and $z \approx 7.59$ (bottom row), capturing the immediate post-starburst moment. From left to right, the panels show DM density, hydrogen number density, gas temperature, and gas metallicity along the same line of sight within $8R_{\rm vir}$ from the center of the galaxy. For reference, we indicate a length scale of 10~kpc at the top of each panel, and also delineate the halo virial radius (dashed circles). After the initial starburst, SN feedback from Pop~III stars (cyan X-symbols) has destroyed the dense gas structure at the center. The strong outflows driven by intense SN feedback consequently expel most of the Pop~II stars (magenta triangles) and gas clouds; the SN feedback also rapidly heats up and enriches nearby gas clouds.}
    \label{fig:projection_plot}
\end{figure*}
\par
Due to the higher $\bar{J}_{21, 0}$, the first star formation is delayed even further in the \textbf{Intermediate} case (LW100E100, green lines), experiencing the initial starburst at $z \approx 7.9$ when the host has grown to $M_{\rm vir} \approx 7.9 \times 10^7 \msun$. Given this additional delay ($\Delta t \sim 20 \rm \,Myr$ compared to the LW10E100 run) in triggering star formation, the simulated galaxy in the \textbf{Intermediate} case can retain a larger amount of cold dense gas than the \textbf{Low} group, resulting in a larger Pop~III stellar mass of $M_{\star} \approx 1.2 \times 10^4 \msun$.
After a short quenching period ($\Delta t \lesssim 15 \rm \,Myr$), the simulated galaxy evolves with a constant star formation history and rapidly transitions to a Pop~II mode, exceeding $M_{\star} \approx 2.1 \times 10^4 \msun$ at the end of the simulation.
\par
 
Finally, the \textbf{High} group exhibits the most delayed first star formation, which does not occur until $z\lesssim 7.69$. Moreover, the halos in this group experience both the largest Pop III starbursts and a rapid transition to Pop~II star formation in the middle of the burst.
During the initial phase of the starburst, both runs form a Pop~III mass of $M_{\star, \rm Pop~III} \geq10^5\msun$ within a short timescale ($\Delta t_\mathrm{burst} < 5\rm \, Myr$). 
As the first Pop~III stars end their lives as SNe, a large amount of cold dense gas survives despite the intense feedback, and Pop~II stars are formed from the enriched gas, finally reaching a total stellar mass $M_{\star} \gtrsim 10^6 \msun$ in both runs (consistent with the upper limit derived by \citealt{Visbal2017} for externally irradiated halos). This rapid transition from a ``pure Pop~III'' to a ``hybrid'' Pop~III/II phase immediately after the first SN explosions is also in agreement with the recent results of \citet{Rusta2025}, suggesting that active Pop III signatures can coexist with signatures of metal-enriched star formation.
\par
To showcase the processes dominating the quenching of star formation and the transition from Pop III to Pop II formation in this last scenario, Fig.~\ref{fig:projection_plot} shows spatially resolved properties (DM density, hydrogen number density, gas temperature and gas metallicity) of the simulated galaxy and its environment from the LW1e3E100 run at the time of the initial Pop~III starburst ($z \approx 7.67$), and when the starburst has just finished ($z \approx 7.59$, $\sim10\rm Myr$ after initial Pop~III starburst). The photoionization feedback from the first starburst heats up the gas to $T\sim 10^4 \rm ~K$, which is not sufficient to disrupt the dense gas clouds within the star-forming region. This leads to the survival of cold and dense metal-free gas clouds at the center that can continue forming stars. After Pop~III stars die, the star-formation mode rapidly transitions to Pop~II. The intense SN feedback from the combined Pop~III and Pop~II starbursts destroys the dense structure at the center, and it eventually expels most of the gas from the galaxy, dragging along newly formed Pop~II stars with the outflowing gas. As a result, at the end of the simulations, the oldest Pop~III stars and their remnants remain in the galaxy center, while Pop II stars are kicked out. This trend is reflected in the stellar mass evolution of the \textbf{High} group, showing sharp peaks with a decreasing trend after the starburst phase.
\par
In summary, we find that extreme LW fluxes ($\bar{J}_{21, 0}\geq 10^3$) can delay the first star-formation episode and cause it to happen in a single, massive burst, demonstrating that the maximum mass for the Pop~III starburst converges to $M_{\rm \star, Pop \, III} \lesssim 10^6 \msun$, consistent with some previous studies (e.g., \citealp{Visbal2017}).

\subsection{Conditions for massive Pop~III starbursts}
\label{Sim:Gas}
\begin{figure*}
    \centering
    \includegraphics[width = 170mm]{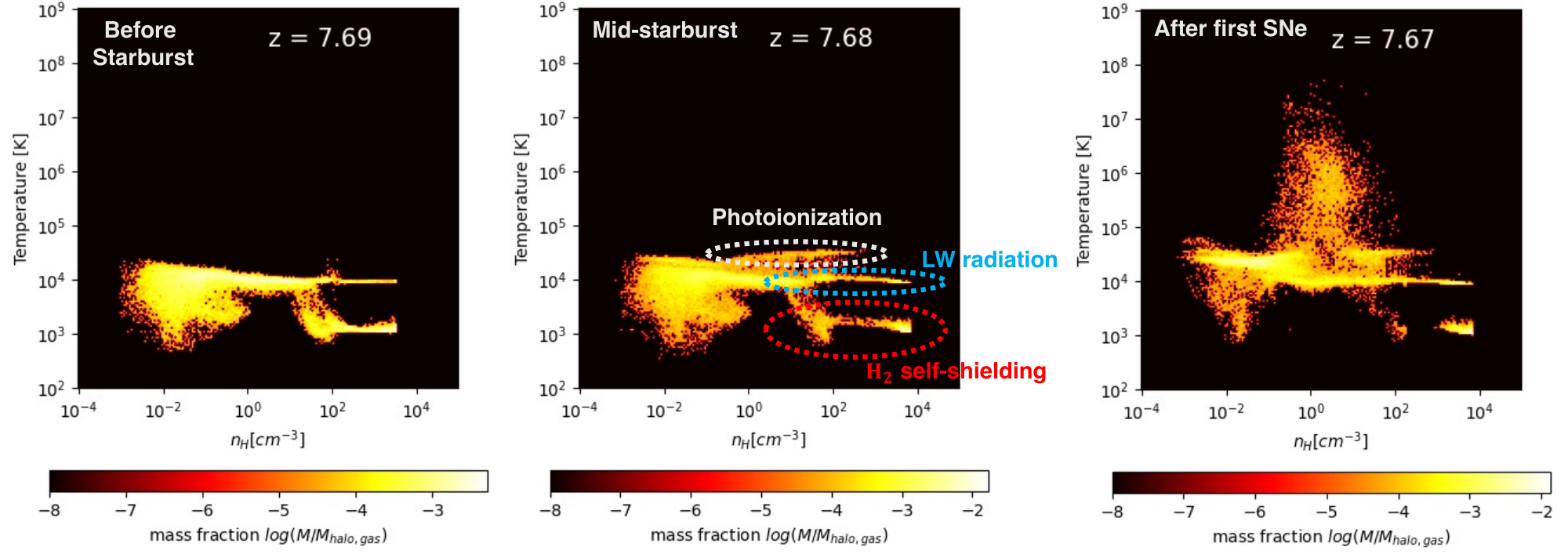}
    \caption{Gas properties of the target galaxy for the LW1e3E100 run before the starburst (left), mid-starburst (middle), and after the first SNe have exploded (right). Before the initial starburst, the $T-n$ phase diagram bifurcates at high densities ($n_{\rm H} \gtrsim 10 \rm \, cm^{-3}$) into a hot track ($T_{\rm gas} \gtrsim 10^4 \rm K$), which is fully exposed to strong LW flux, and a cold one ($T_{\rm gas} < 10^4 \rm K$), where the gas is self-shielded by $\rm H_{2}$. 
    When gas particles reach $n_{\rm H} \gtrsim5\times 10^2 \rm \, cm^{-3}$, a small abundance of $\rm H_{2}$ molecules can trigger the $\rm H_{2}$ self-shielding effect, consequently cooling down a fraction of the dense gas.
    Eventually, intense SN feedback from the starburst has heated nearby gas to $T_{\rm gas} \lesssim 10^8 ~\rm K$, thus destroying the central dense gas structure. Most of the dense gas is expelled through outflows, while second-generation stars are still forming inside the remaining dense gas.}
    \label{fig:Phase-diagram}
\end{figure*}

As discussed above, massive Pop~III starbursts with $M_{\star} \gtrsim 10^5 \msun$ only occur in simulated galaxies of the \textbf{High} group ($\bar{J}_{21,0}\geq 10^3$). To determine the physical reason for this behavior, we next investigate the properties of the central gas clouds within this group. Fig.~\ref{fig:Phase-diagram} shows the phase diagram for the gas within the LW1e3E100 target halo, before the starburst (left), mid-starburst (middle), and after the first SN explosions (right). Prior to the starburst and above $n_\mathrm{H} \sim 10 ~\mathrm{cm^{-3}}$, the gas evolves towards higher density following two separate tracks, one for the hot ($T_{\rm gas} \sim 10^4 \rm \,K$) and one for the cold ($T_{\rm gas} < 10^4 \rm \,K$) phase, while the hot branch is absent in simulations with lower LW flux (see Fig.~\ref{fig:Gas-phase-APX}). The hot gas reaches the threshold density, $n_{\rm H, thr}$, set by our resolution limit, but cannot immediately experience star formation due to its high temperature, allowing additional gas to collapse into the deepening potential well.
\par
When the central gas clouds reach $n_{\rm } \gtrsim 500 \cmci$, the self-shielding against the LW radiation from a small fraction of hydrogen molecules ($f_{\rm H_{2}} \lesssim 10^{-6}$) finally enables efficient $\rm H_2$ formation, thus activating the molecular-cooling channel.
While the cold phase remains confined within the core of the simulated galaxy, an extended layer of hot gas surrounds the cold core in a Matryoshka-like structure (see Fig.~\ref{fig:sliceplot}).
This morphology is a consequence of our idealized modeling of the LW radiation, in which we assume a uniform distribution of sources, together with a prescription for self-shielding that depends on the inhomogeneous density field.  We will further discuss the combined effects of these prescriptions in Section~\ref{Caveats}.
\par
After the cold phase has emerged, it continues to grow in mass until the density reaches $n_{\rm H} \gtrsim 3\times 10^3 \cmci$, where the gas clouds efficiently cool to $T_{\rm gas} \leq 10^3 \rm \,K$ and activates star formation. As a result, star-forming gas in the \textbf{High} group reaches the highest densities among our simulated sample, corresponding to the shortest $\tau_{\star}$ values, compared with the lower $\bar{J}_{21,0}$ simulations. This leads to an intense Pop~III starburst, forming $M_{\star} \gtrsim 10^5 \msun$ within $\lesssim 3 \rm Myr$. 
During the starburst (middle panel of Fig.~\ref{fig:Phase-diagram}), we can discern the emergence of another phase (white dashed circle) with higher temperature due to photoionization feedback. The photoionization feedback from newly born Pop~III stars starts to disrupt the pre-starburst ``Matryoshka-like'' structure of shielded, cold gas embedded in the hotter envelope exposed to the full LW flux. However, the hot, photo-dissociated (blue dashed circle) and cold, shielded (red dashed circle) phases remain, and star formation still proceeds in the cold phase.
After the short lifetime of massive Pop~III stars, the first SN explosions occur (right panel of Fig.~\ref{fig:Phase-diagram}), evacuating most of the gas in the hot and cold branches and finally quenching star formation. 

\par
Why do the galaxies in the \textbf{High} group attain the largest stellar mass during the initial Pop~III starburst, although they are exposed to the most intense LW flux ($\bar{J}_{21,0} \geq 10^3$)? We find that in simulated galaxies with lower LW flux levels, the star-formation criterion is satisfied in gas clouds at $n_{\rm H} \lesssim 500\cmci$, whereas the gas within the \textbf{High} group reaches much higher densities of $n_{\rm H} \gtrsim 3\times 10^3\cmci$ before it becomes cool enough to allow star formation.
Our simulations thus point towards two key factors that determine the mass of the Pop~III starburst: its delayed onset and the higher gas densities involved. Due to the delayed onset of star formation, the simulated galaxies can accrete more baryonic material, resulting in a larger amount of dense gas accumulated at the center. For example, the simulated galaxy in the moderate LW-flux case (LW10E100) contains dense gas clouds with $M_{\rm >n_{\rm H, thr}} \approx 10^3 \msun$ before first star formation, which is almost 3 orders of magnitude lower than for the LW1e3E100 run. Moreover, star formation within galaxies of the \textbf{High} group proceeds at relatively higher gas density compared with lower LW-flux simulations, which leads to shorter $\tau_{\star}$, so that more mass can be converted into stars before SN feedback kicks in and quenches star formation. In conclusion, these two key factors yield intense Pop~III star formation bursts, reminiscent of aspects of the ``feedback-free'' starburst scenario proposed by \citet{Dekel2023}.

\subsection{Metallicity evolution}
\label{Sim:Met}
\begin{figure}
    \centering
    \includegraphics[width=85mm]{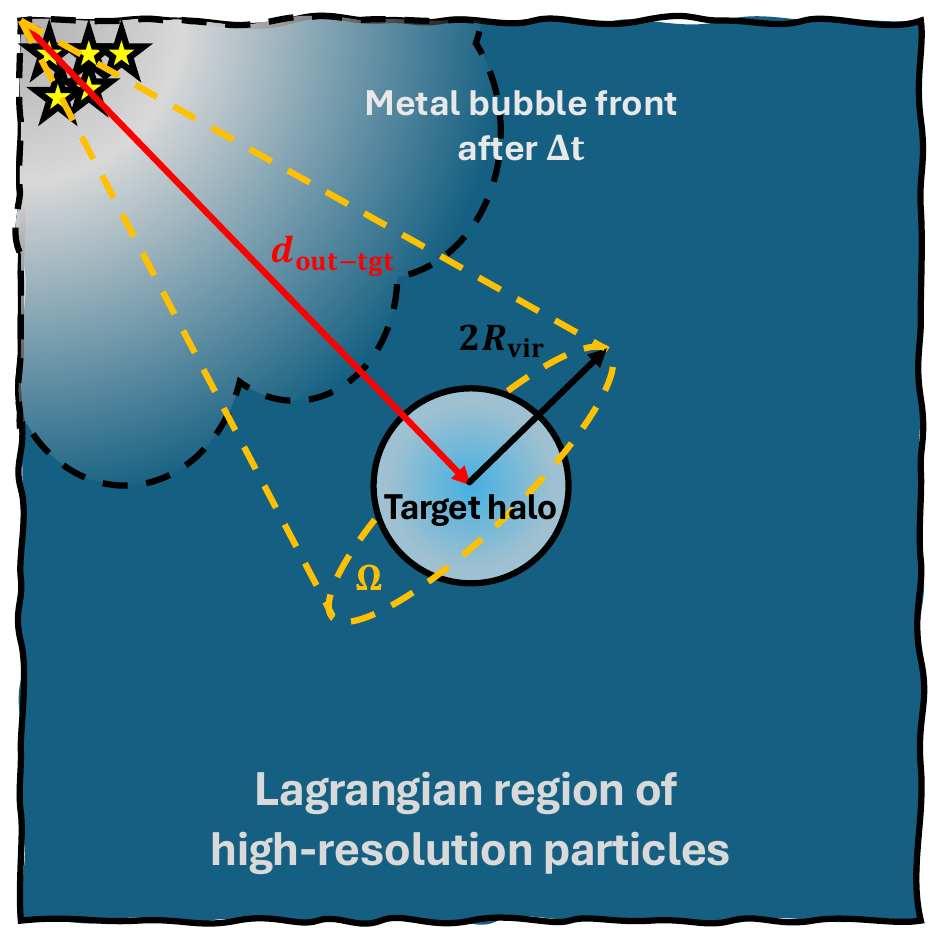}
    \caption{Illustrating the modeling of external metal pollution. Here, we show the Lagrangian region of high-resolution particles, containing the target halo for Pop~III star formation at its center (sky blue circle). The star symbols represent star particles external to the target halo (out-stars), which are the sources of the external metal enrichment. We mark the extent of the metal bubble (black dashed line), reached at $\Delta t$ after the start of the enrichment process. The red arrow ($d_{\rm out-tgt}$) shows the distance between the out-stars host halo and the center of the target halo. The cone (yellow dashed line) indicates the solid angle ($\Omega$) considered for tracking the metal bubble front, covering an area of $2R_{\rm vir}$ at the distance $d_{\rm out-tgt}$.}
    \label{fig:external_metal_cartoon}
\end{figure}
\par
We next discuss the metallicity evolution in our simulations, specifically the evolution of the metal bubble front that can externally enrich our target galaxies and of the resulting gas-phase metallicity. A wide range of values has been proposed for the threshold metallicity, $Z_{\rm thr}$, that guides the transition from Pop~III to Pop~II star formation, reflecting the complexity of the underlying physical processes \citep[e.g.,][]{Karlsson2013}, such as dust-induced fragmentation (e.g., \citealp{Omukai2005}). Moreover, several theoretical studies suggested evolving IMFs that gradually shift their character from top-heavy to bottom-heavy, depending on the temperature of the CMB and on gas metallicity (e.g., \citealp{Chon2022}). Nevertheless, there is general agreement on $Z_{\rm thr} \lesssim 10^{-3} \Zsun$, implying that Pop~III star formation could be terminated in the presence of even moderate metal enrichment (e.g., \citealp{Magg2022, Katz2023, Yamaguchi2023, Ventura2024MNRAS, Brauer2025}).
For example, using cosmological simulations, \citet{Jeon_2017} found that low-metallicity (Pop~II) stars were able to form within their targeted dwarf galaxies at $z \gtrsim 6$ because of external metal enrichment from neighboring halos. 
Even though star formation in the progenitor minihalos is suppressed by the high LW flux, an infall of metals from nearby systems can still pollute our target halo before internal metal enrichment is triggered.
Therefore, it is crucial to investigate the evolution of the metal bubbles generated by nearby systems.
\par
\begin{figure}
    \centering
    \includegraphics[width=85mm]{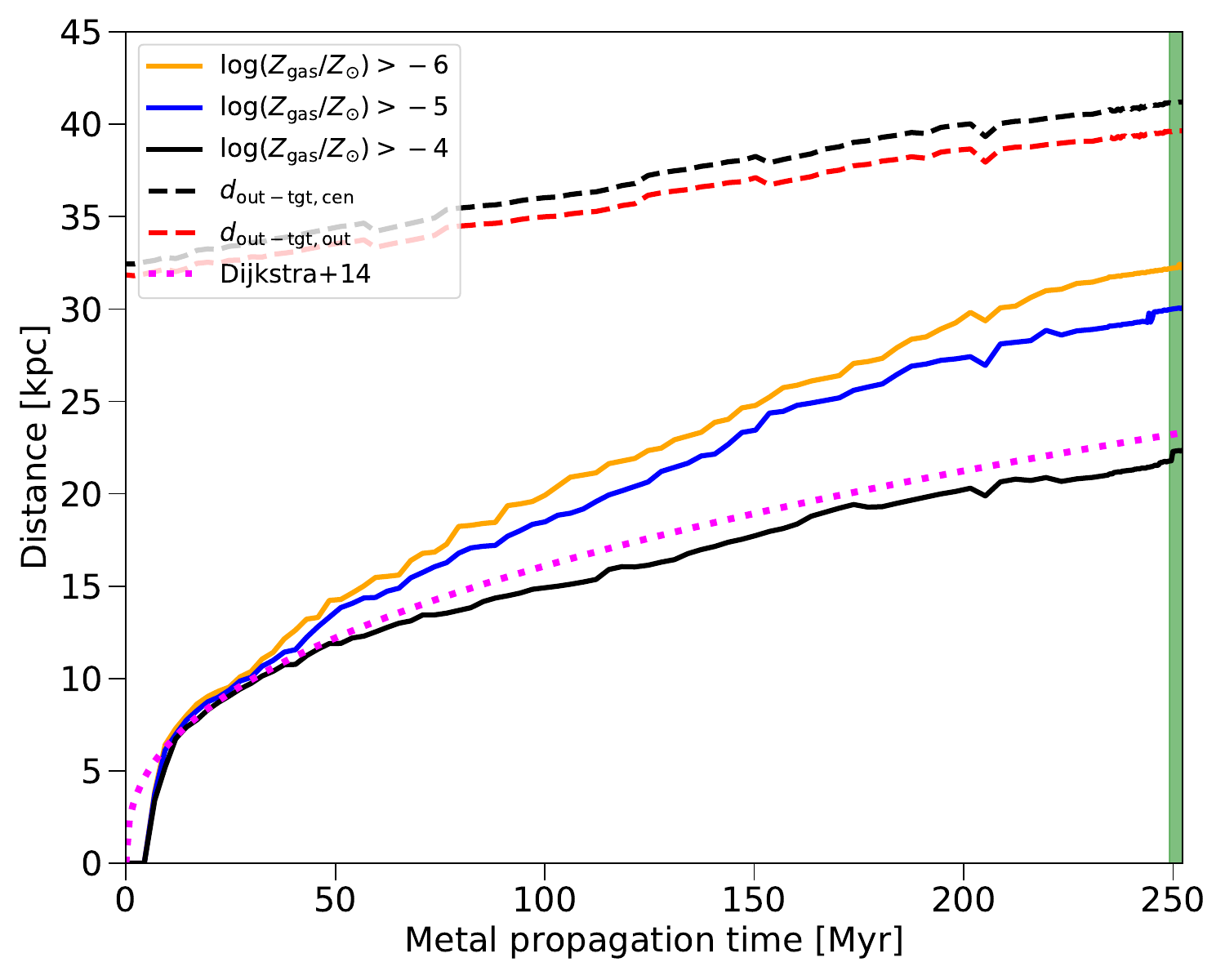}
    \caption{Distance from the out-stars host halo to the metal bubble front as a function of time, followed until internal enrichment is triggered within our simulated galaxy in order to constrain the external metal enrichment process. Each solid colored line shows the evolution of the bubble front for select metallicities: $\log(Z_{\rm gas}/\Zsun) > -6$ (orange), $\log(Z_{\rm gas}/\Zsun) > -5$ (blue), and $\log(Z_{\rm gas}/\Zsun) > -4$ (black). The dashed lines show the distance to the center of the target halo (black) and to its outer boundary (red). For comparison, we show the analytic result using the galactic outflow model of \citet[][dotted magenta line]{Dijkstra2014}. 
    We also indicate the time when the Pop~III starburst occurs within our simulated galaxy (marked as a green shaded region), demonstrating that metal bubbles from nearby enrichment events cannot reach the target halo on time to prevent metal-free star formation.}
    \label{fig:metal_bubble}
\end{figure}
\par
To investigate the impact of external metal enrichment, we track the expansion of metal bubbles from sources within the zoom-in region that can significantly influence our simulated galaxy (see the illustration in Fig.~\ref{fig:external_metal_cartoon}), focusing on the closest metal-producing system, which experiences an intense starburst of $M_{\star} \approx 1.6 \times10^5 \msun$ within a $M_{\rm vir} \sim 10^8\msun$ host halo. We specifically track the propagation of the bubble front for three threshold metallicities, $Z_{\rm b, thr}$: $\log(Z_{\rm gas}/Z_{\star})>-6$, $\log(Z_{\rm gas}/Z_{\star})>-5$, and $\log(Z_{\rm gas}/Z_{\star})>-4$, only considering gas particles within the solid angle $\Omega$ (marked with a yellow dashed line in Fig.~\ref{fig:external_metal_cartoon}) that covers an area of $\pi (2R_{\rm vir})^2$ at the distance between the center of the out-star halo and the simulated galaxy, $d_{\rm out-tgt}$. We expect this $\Omega$ to be sufficiently large for tracking all gas particles that could influence our simulated galaxies. As a final step, we compute the time evolution of the maximum distance from the center of the out-stars host halo to the metal bubble fronts, until the first internal metal enrichment inside our simulated galaxies is triggered.
\par
Fig.~\ref{fig:metal_bubble} shows the propagation of the metal bubble front from the out-star host halo in the LW1e3E100 run as a function of time. Each solid line exhibits the distance from the center of the out-star host halo to the metal bubble front for a specific threshold metallicity, compared to the evolving separation between target and out-star host halo, center to center (black dashed line) and center to the boundary of the simulated galaxy (red dashed line). Furthermore, we reproduce the analytic outflow solution from \citet{Dijkstra2014}, using their equation~(4) and assuming an IGM number density of $n = \Omega_{\rm b} \rho_{\rm crit}/m_{\rm p}$, marked with the magenta dotted line. As can be seen, the simulation results for the
propagation of the metal front with $\log(Z_{\rm gas}/\zsun) > -4$ are well matched by the analytic results from \citet{Dijkstra2014}. Furthermore, it is evident that the metal-enriched bubble cannot reach the boundary of the target halo before the first starburst occurs in our simulated galaxy. At the onset of internal metal enrichment, even the metal bubble with $\log(Z_{\rm gas}/\zsun) > -6$ cannot reach our simulated galaxy, thus rendering any external enrichment unimportant here\footnote{There would be an even further delay of $\Delta t_{\rm diff} \approx 50 \rm \, Myr$ before metals deposited at the boundary of the target halo could reach its center via turbulent diffusion \citep[][their equations~(9) and (10) with a characteristic length of $l_{\rm turb} = R_{\rm vir}$]{Ji2014}.}.

\begin{figure}
    \centering
    \includegraphics[width=85mm]{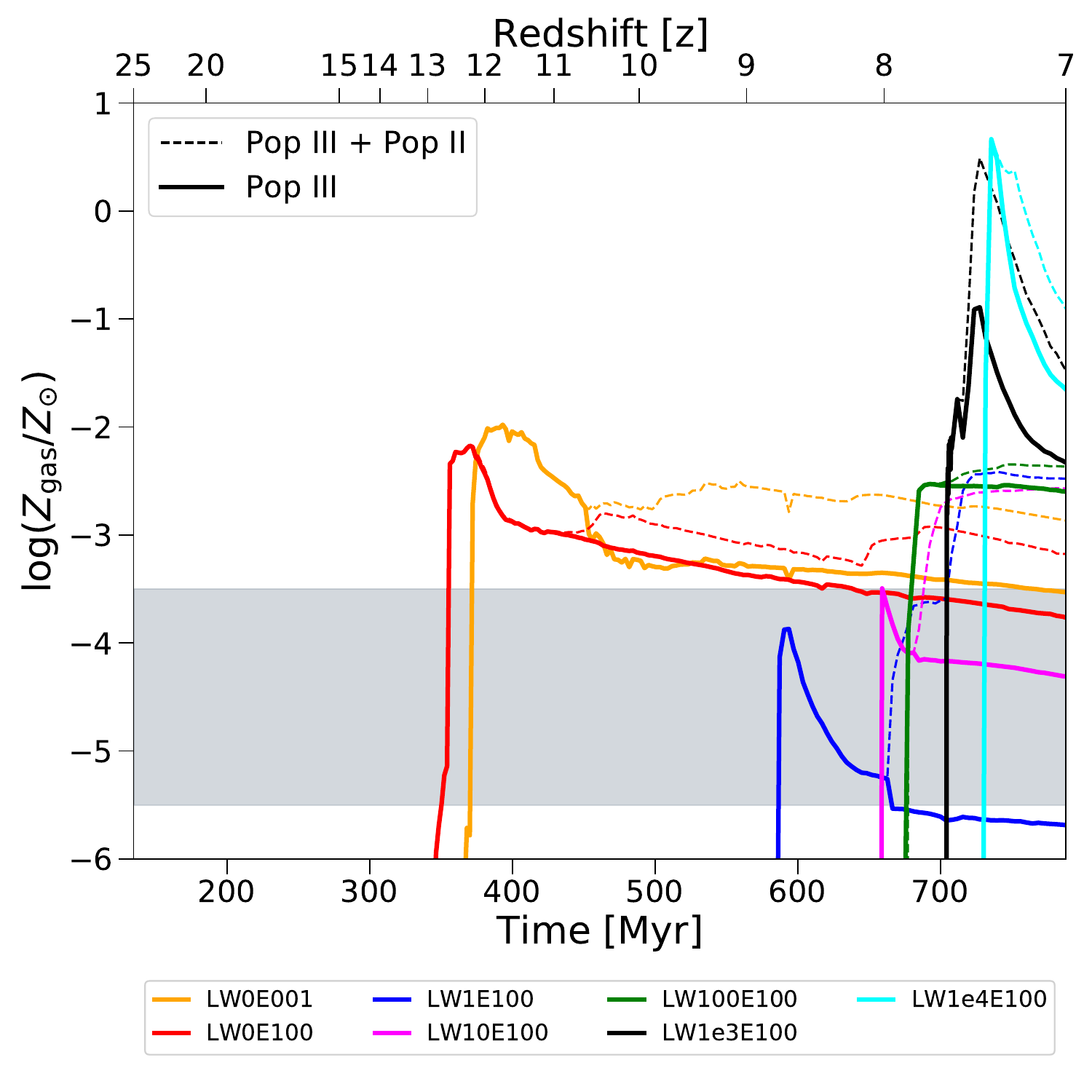}
    \caption{Evolution of the gas-phase metallicity within the virial radius of our simulated galaxies as a function of time and redshift. The dashed lines show the gas-phase metallicity contributed by the total amount of metals produced from both Pop~III and Pop~II stars, while the solid lines show the metallicity with the metals produced by Pop~III stars only. We employ the same convention for the lines as in Fig.~\ref{fig:mass_evolution}, and the gray shaded region indicates the range of threshold metallicities, from $\log Z_{\rm thr} = -5.5$ to $-3.5$. 
    }
    \label{fig:gas_phase_metallicity_evolution}
\end{figure}
\par
Fig.~\ref{fig:gas_phase_metallicity_evolution} summarizes the gas-phase metallicity evolution within the virial radius of our target galaxy as a function of cosmic time and redshift. We separately show metals produced from Pop~III stars and from total stellar activity (Pop~III and Pop~II), with solid and dashed lines, respectively. 
The gas-phase metallicity of the target galaxy already exceeds our assumed $Z_{\rm thr} = 10^{-5.5} ~\mathrm{Z_\odot}$ after the first episode of metal enrichment in all simulations. Consequently, metal enrichment is thereafter dominated by Pop~II stars. After the first star-formation episode, the gas-phase metallicity rapidly declines, which is a consequence of metal-rich SN outflows removing metals from the galaxy. However, this rapid decrease is mitigated in the \textbf{Intermediate} group (LW100E100), as the galaxies in this simulation set experience a constant star formation history after the first starburst with only a short quenching period, which replenishes the ejected metal content.
As a consequence of the intense Pop~III starburst ($M_{\rm \star, Pop~III} \gtrsim 10^5\msun$) in the most extreme LW-flux cases, metals produced by Pop~III stars rapidly enrich the surrounding ISM and eventually reach $\log (Z_{\rm gas}/\Zsun) \gtrsim -1$. This behavior is consistent with previous results from \citet{Rusta2025}, suggesting that galaxies with a significant active Pop III component can be observed even in the presence of metal-pollution from self-enrichment.
\par
Finally, we consider the sensitivity of our results on the value for $Z_{\rm thr}$. 
Even increasing the threshold value to $\log (Z_{\rm thr}) = -3.5$ (the upper boundary of the gray shaded region in Fig.~\ref{fig:gas_phase_metallicity_evolution}), we conclude that there will be no critical impact on our simulation results, as most of the simulated galaxies exceed any choice of $Z_{\rm thr}$ during the first starburst phase with rapid SN metal enrichment from the first (Pop~III) stars.

\section{Observational signatures}
\subsection{Post-processing results}
\label{Result:Obs}

\begin{figure*}
    \centering
    \includegraphics[width=170mm]{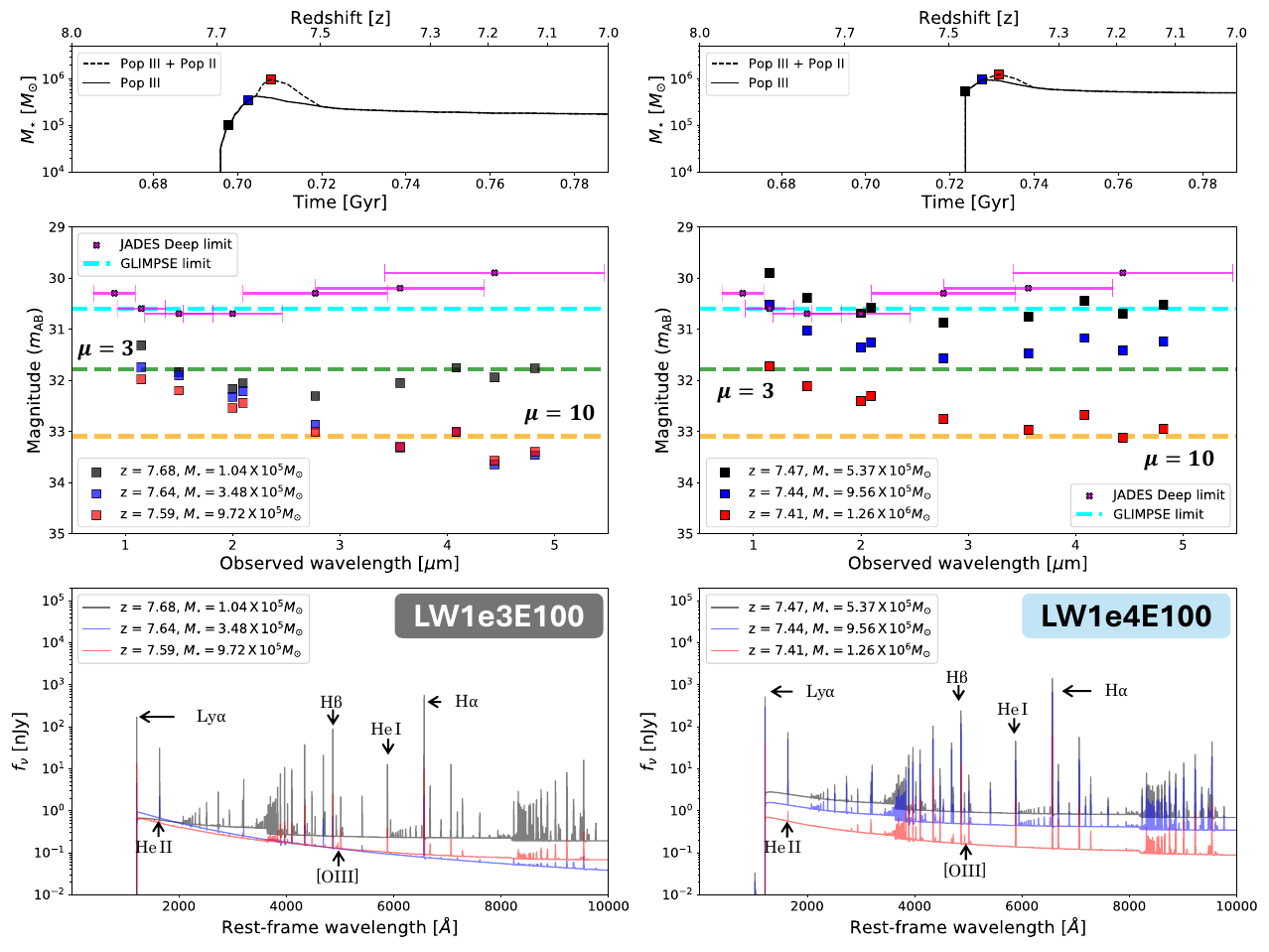}
    \caption{The post-processed SEDs and expected AB magnitudes for select JWST filters, based on our simulated galaxies during a starburst. The left panels correspond to the LW1e3E100 run, and the right panels to the LW1e4E100 run. 
    While the top row shows the stellar mass evolution, marking three illustrative phases with colored square symbols, panels in the middle row show AB magnitudes calculated for select JWST NIRCam filters (square symbols) at the observed frame, including wide and medium band ones.
    The magenta x-shaped symbols with error bars represent the limiting magnitude of the JADES Deep survey (\citealp{Eisenstein2023}), together with their respective bandwidth, and the dashed cyan lines show the limiting magnitude of the GLIMPSE survey (\citealp{atek2025GLIMPSE}). We also indicate effective limiting magnitudes for the GLIMPSE survey, assuming gravitational lensing magnifications of $\mu = 3$ (green dashed line) and $\mu = 10$ (orange dashed line). Panels in the bottom row show rest-frame SEDs, in units of nJy, for three characteristic phases: the initial Pop~III starburst (black), the end of this phase before transition into Pop~II star formation (blue), and the second starburst phase consisting of Pop~II stars (red). 
    }
    \label{fig:observation}
\end{figure*}

Next, we examine the observability of our simulated galaxies. Due to the computational cost, we only perform post-processing to derive mock spectra for galaxies in the \textbf{High} group (LW1e3E100, LW1e4E100), which have been found to produce the maximum Pop~III starbursts ($M_{\star, \rm Pop~III} >10^5\msun$). Fig.~\ref{fig:observation} shows the post-processing results during three different stages of the starburst for the LW1e3E100 (left) and LW1e4E100 run (right), respectively. 
We show limiting magnitudes for two select JWST surveys, JADES Deep (\citealp{Eisenstein2023}, magenta x symbols) and GLIMPSE (\citealp{atek2025GLIMPSE}, cyan dashed line) as a reference in the middle panels. To account for gravitational lensing, we introduce an ``effective limiting magnitude'' for the GLIMPSE survey, following the functional form
\begin{equation}
    m_{\rm AB, \, eff} = m_{\rm AB}+2.5\log(\mu),
\end{equation}
where $m_{\rm AB, \, eff}$ is the effective limiting magnitude for a given magnification factor ($\mu$), and $m_{\rm AB}$ is the authentic limiting magnitude of the survey.
For illustration, we consider two cases, for magnification factors of $\mu = 3$ (green dashed line) and $\mu = 10$ (orange dashed line).
\par
For the LW1e3E100 run, we find that, despite having experienced a large Pop~III starburst with $M_{\rm \star, Pop \, III} \gtrsim 10^5 \msun$, resulting magnitudes are still $1-2 \, \rm mag$ lower than the limiting magnitudes of both the JADES Deep and GLIMPSE survey. Even considering the sensitivity reached with the NGDEEP+MIDIS survey ($m_{\rm AB}\sim 31.1$; G. Leung et al., in prep.), our predicted magnitudes are still too faint for detection, indicating the crucial role of gravitational lensing.  
After the most massive Pop~III stars formed in the starburst explode as SNe, they trigger a transition to Pop~II star formation. As a result, the magnitudes in each JWST filter evolve towards fainter values, despite the increasing stellar mass.
\par
The magnitudes of the LW1e4E100 run show a similar evolution, with a decreasing trend following the first Pop~III starburst. 
However, in contrast to the LW1e3E100 run, magnitudes during the initial Pop~III starburst exceed the limiting magnitude of both surveys in specific filters, e.g., F115W (F115W and F277W) for JADES (GLIMPSE), due to the higher continuum level and emission line fluxes from a more vigorous Pop~III starburst. This indicates promising observability even within surveys without the support of gravitational lensing.
\par
Gravitational lensing can significantly improve detectability by boosting the luminosity of our simulated galaxies. With moderate lensing ($\mu =3$), the resulting magnitudes for the shortest-wavelength (F115W, F150W) filters exceed the limiting values during the initial phase of the Pop~III starburst for the LW1e3E100 run, and over the entire Pop~III starburst for the LW1e4E100 case. Moreover, both runs exceed the limiting magnitude, if a high magnification of $\mu = 10$ is assumed, consequently allowing detection through the entire starburst phase.
\par
During the initial Pop~III starburst, hard ionizing photons from massive stars produce a significant He~II ~1640\,\AA\ emission line,  with $\log \rm (He~II/H\alpha) \approx -0.7$, ${\rm EW}(\rm He~II)\gtrsim76$~\AA\, for both runs, together with strong nebular continuum features. Both runs exhibit typical Pop~III signatures until the end of the Pop~III starburst phase, accompanied by an absence of metal lines such as the [O~III] doublet. Afterwards, as the simulations quickly transition to Pop~II star formation (within $\lesssim1\rm Myr$), the SEDs of both runs show a significant decrease in the He~II 1640\,\AA\, flux, such that $\log \rm (He~II/H\alpha) \approx -1.43 \, (-1.62)$ for LW1e3E100 (LW1e4E100), together with the appearance of metal lines (see Table~\ref{tab:line_luminosity} for details).
\par
\begin{figure}
    \centering
    \includegraphics[width=85mm]{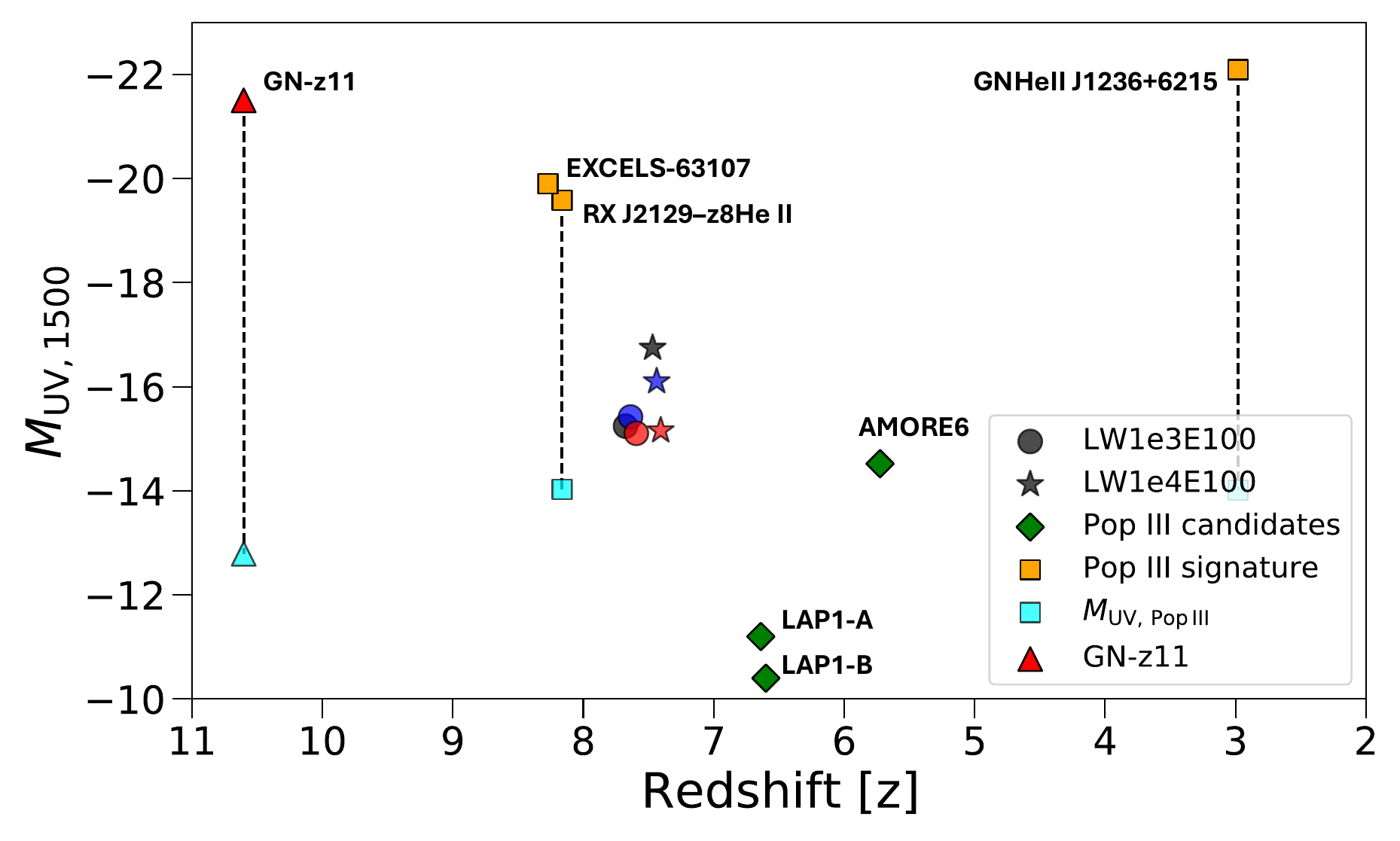}
    \caption{$M_{\rm UV, 1500}$ evolution of simulated galaxies in \textbf{High} group as a function of redshift. The circle and star symbols indicate the LW1e3E100 and LW1e4E100 set, with colors indicating the same evolutionary phases as in Fig.~\ref{fig:observation}. 
    For comparison, we show observed pure Pop~III galaxy candidates (LAP1-A (\citealp{Vanzella2023}), LAP1-B (\citealp{Nakajima2025}), and AMORE6 (\citealp{Morishita2025})), as well as galaxies including potential Pop~III signatures (RX~J2129–z8He~II (\citealp{Wang2024}), EXCELS-63107 (\citealp{Cullen2025}), and GNHeII~J1236+6215 (\citealp{Mondal2025}), marked as green diamonds and orange squares with their IDs, respectively. We represent GN-z11 (\citealp{Bunker2023}), containing a He~II clump nearby, with a red triangle. We also indicate (with cyan symbols) the $M_{\rm UV}$ inferred for Pop~III stars within larger host systems, such as GN-z11. 
    }
    \label{fig:UVmag}
\end{figure}

Finally, we show the evolution of the absolute UV magnitude at 1500~\AA~($M_{\rm UV, 1500}$) of simulated galaxies in the \textbf{High} group as a function of redshift in Fig.~\ref{fig:UVmag}, computed by adopting a boxcar filter centered on $1500$~\AA\, with a bandwidth of $\Delta\lambda = 200$~\AA, avoiding the contribution of He~II 1640~\AA. For comparison, we also illustrate the $M_{\rm UV, 1500}$ of pure Pop~III galaxy candidates (green diamonds, \citealp{Vanzella2023, Morishita2025, Nakajima2025}), as well as for candidates where a potential Pop~III contribution has been suggested to explain their hard spectral signatures (orange squares, \citealp{Wang2024, Cullen2025, Mondal2025}), observed in JWST surveys. We also mark the $M_{\rm UV, 1500}$ of GN-z11 (\citealp{Bunker2023}), in which a He~II-emitting clump compatible with Pop~III stellar populations has recently been observed in the vicinity of the luminous source at the center (\citealp{Maiolino2024b}).
\par
Due to the higher Pop~III stellar mass, the simulated galaxy in the LW1e4E100 run exhibits a rapid evolution from $M_{\rm UV, 1500} \approx -16.7$ to $ \approx -15.2$. On the other hand, the dominant Pop~II stellar component in the LW1e3E100 case maintains an almost constant UV flux throughout the starburst phase, in the range of $-15.5 \lesssim M_{\rm UV, 1500} \lesssim -15.0$. 
The UV magnitudes inferred from our simulations are largely similar to select Pop~III candidates in the EoR, such as AMORE6 (\citealp{Morishita2025}).
However, our post-processed $M_{\rm UV, 1500}$ values are significantly fainter than those for observed galaxies that include a potential Pop~III signature.
This large discrepancy in $M_{\rm UV, 1500}$ may reflect the sub-dominant Pop~III contribution in galaxy candidates with a potential Pop~III signature consisting of a mixed population of Pop~III and Pop~II stars. For instance, \citet{Mondal2025} and \citet{Wang2024} indicated that their galaxy candidates could plausibly host small pockets of Pop~III stars with total Pop~III stellar mass $M_{\star, \rm Pop~III} \sim 7.7\times 10^5 \msun$ and $M_{\star. \rm Pop~III} \approx 7.8 \pm 1.4 \times 10^5 \msun$, respectively, comparable to the results of our \textbf{High} simulation group. Based on their estimated Pop~III stellar masses, we compute the corresponding $M_{\rm UV, 1500}$ expected for these Pop~III systems, employing SEDs evaluated at the zero-age main sequence (ZAMS). We find $M_{\rm UV, Pop~III} \lesssim -14$, consistent with our derived upper limit for Pop~III starbursts.

\subsection{Finding Pop~III galaxies}
\label{Res:Statics}
In this work, we only focused on a single galaxy and its environment, using the zoom-in technique to reach high resolution within the cosmological region of interest. This allowed us to explore a wide parameter space for the strength of the LW background, at controlled computational cost. On the other hand, we could not derive the overall abundance of similar systems directly from our simulation or post-processed outputs. Statistical properties of this galaxy population, such as its contribution to the overall luminosity function of Pop~III galaxies or to the Pop~III cosmic star formation rate density, play nonetheless a crucial role in determining how Pop~III starbursts shape the evolution of the Universe, as well as the volume that future surveys need to probe in order to detect them.
\par
Therefore, we estimate the number of detectable Pop~III galaxies in current JWST surveys through a simple analytic argument. Using the halo mass function from \citet{Tinker2008}, we obtain the average number density of halos close to the virial mass of our simulated galaxies ($8\times10^7\msun \lesssim M_{\rm vir} \lesssim 2 \times10^8\msun$) at $6 \leq z \leq 8$, giving $n_{\rm ACH} \sim 8.3 \rm \, cMpc^{-3}$. The expected number of pristine halos experiencing strong LW radiation ($\bar{J}_{21,0} \geq 10^3$) then is:
\begin{equation}
    N = n \times V = n_{\rm ACH}\times f_{\bar{J}_{21,0}\geq 10^3} \times \frac{\Delta t_{\rm Pop~III}}{\tau} \times (1-Q(>Z_{\rm thr})) \times V,
\end{equation}
where $f_{\bar{J}_{21,0}\geq 10^3}$ is the probability that an ACH is exposed to LW flux with $\bar{J}_{21,0} \geq 10^3 $, $Q(>Z_{\rm thr})$ the cosmic volume fraction of gas at a metallicity higher than $Z_{\rm thr}$, $\Delta t_{\rm Pop~III} = 5\rm \, Myr$ for the Pop~III starburst timescale, $\tau$ is the time interval corresponding to $6<z<8$, and $V$ the effective comoving volume of the considered surveys.
\par
We adopt $f_{\bar{J}_{21,0}\geq 10^3} = 2.9\times 10^{-3}$, integrating our fit to the analytic solution of \citet{Dijkstra2008} at $z=10$ (the dotted line in their top panel of figure~6) over the range $10^3 \leq J_{\rm LW} \leq 10^5$.
We cautiously note that this is a conservative choice, since the number of halos experiencing high LW radiation is expected to increase during the EoR (e.g., \citealp{Ahn2009, Agarwal2012}). We further assume that $1-Q_{\rm thr} \approx 0.99$, adopting the results from \citet{Pallottini2014} (with $Z_{\rm thr} = 10^{-8}\zsun$), averaged over $6 \leq z \leq 8$. 
As a result, the expected number density of pristine or quasi-pristine ACHs that experience environmental conditions similar to those producing Pop~III starbursts in our simulations (our \textbf{High} group) is $n \approx4 \times 10^{-4} \rm \,cMpc^{-3}$.
\par
Considering the volume covered by GLIMPSE, for an effective source-plane area of $4.4\rm \, arcmin^2$  at $z\sim6$ \citep{atek2025GLIMPSE}, we find an expected number of Pop~III starbursts of $N_{\rm GLIMPSE} \sim 9$. We cautiously note that here the survey volume of GLIMPSE includes both lensed and unlensed fields.
We also consider large surveys that do not involve gravitational lensing, such as JADES (\citealp{Eisenstein2023}, $167 \rm \, arcmin^2$). Since only post-processed results from the LW1e4E100 run exceed the JADES limiting magnitude, we now adopt $f_{\bar{J}_{21} \geq10^4}$, restricting our predictions to $\bar{J}_{21,0} \geq 10^4$ scenarios. 
Under this assumption, the number density of halos experiencing conditions similar to our LW1e4E100 run is $n_{\rm \bar{J}_{21,0}\geq 10^4} \approx1.25\times 10^{-5} \rm \, cMpc^{-3}$. Adopting the JADES survey volume, we estimate a number of detectable Pop~III galaxies ($m_{\rm AB} \lesssim 31$) of $N_{\rm JADES} \sim 1.25$.
\par
These pessimistic predictions for the number of detectable Pop~III galaxies stem from the two strict conditions imposed on their formation sites -- namely a high LW intensity from nearby sources, and the presence of metal-free pockets. On the other hand, these two conditions might provide leverage for the selection of promising candidates when searching for bright Pop~III systems. In fact, to enable efficient Pop~III formation, atomic-cooling halos need to experience strong LW radiation in the absence of metal pollution, implying that Pop~III star clusters might form within pristine pockets, which are close enough to massive, star-forming galaxies to be influenced by their radiative output (e.g., \citealp{Venditti2024, Zier2025, Venditti2026}). The survival of these pristine pockets is plausible, as the typical timescale for metal propagation is longer than the propagation timescale of photons (e.g., \citealp{Pallottini2014}). 
\par
For example, \citet{Trinca2026} found that DCBHs (which are an alternative outcome proposed for metal-free gas in strongly irradiated regions, also see our discussion in Section~\ref{Caveats}) are formed mostly within pristine pockets near the massive, quasar-hosting system in their simulation\footnote{Note that, although their critical LW threshold for DCBH formation is $J_{\rm crit} = 300J_{21}$, this criterion roughly matches the favored condition for Pop~III starbursts in our simulations.}. This suggests that promising formation sites for either massive Pop~III starbursts or DCBHs 
may be found within overdense regions of the cosmic web, possibly concealed by the strong fingerprints of massive star-forming (or even quasar-hosting) systems in the environment. \citet{Venditti2023} also proposed a similar conclusion, arguing that Pop~III stars can be found in the outskirts of metal-enriched regions, within massive halos located in high-density cosmic patches. Therefore, we conclude that detecting Pop~III galaxies is challenging, but plausible in both lensing or large-volume surveys, and that ideal conditions for Pop~III starbursts may be encountered in pristine pockets near massive, star-forming systems at late times, providing precious indications for the selection of promising targets.

\section{Discussion}
\label{Caveats}
\subsection{Caveats and limitations}
Although we are focusing on a scenario of locally high LW radiation feedback produced and propagated from specific nearby sources, we adopt a uniform LW emission, similar to how a uniform external UV background is often implemented \citep[e.g.,][]{Haardt2012}. The LW flux, as realized in our simulations, thus shows isotropic and homogeneous features, only depending on the local $\rm H_{2}$ self-shielding of each gas particle. As described in Section~\ref{Sim:Gas}, this might significantly affect the distribution and morphology of dense gas within the target halo, including the Matryoshka doll-like structure of the bifurcated, dense gas phase in Fig.~\ref{fig:Phase-diagram} \citep[for a more realistic treatment, see, e.g.,][]{Regan2014}. 
\par
It is important to note that the higher values of $\bar{J}_{21,0}$ considered in our simulations straddle the critical intensity commonly assumed to separate delayed star formation and DCBH formation scenarios (see, e.g., \citealp{Shang2010, Dijkstra2014, Regan2014, Wise2019, Prole2024, Jeon2025a, Jeon2025b, Jeon2025, Trinca2026}). Furthermore, \citet{Wise2019} suggested that SMSs could form even within halos with a LW flux of $J_{\rm LW} = 3J_{21}$. 
\citet{Regan2014} also investigated the formation of DCBHs in atomic-cooling halos, suggesting that $J_{\rm LW} \gtrsim 10^3J_{21}$ is required to form a DCBH with a mass of $M_{\rm BH}\approx10^5 \msun$. 
Hence, the implementation of sub-grid recipes for the SMS or DCBH scenarios may lead to the formation of a DCBH with a mass $M_{\rm BH} \lesssim 10^6 \msun$ within galaxies in our \textbf{High} group.
\par
The formation of a DCBH inside the halo can negatively affect the gas capability to form stars, due to its feedback and accretion processes \citep[e.g.,][]{Jeon2025a}. On the other hand, positive feedback around the DCBH could instead trigger Pop~III starbursts \citep[e.g.,][]{Aykutalp2020}. Despite the presence of a strong LW background ($J_{\rm LW} = 10^3J_{21}$), X-ray feedback from the DCBH can enhance the formation of $\rm H_{2}$, and consequently decrease the cooling time of nearby gas clouds. 
This suggests that DCBH and Pop~III starbursts could coexist in high-mass halos, although the final mass of the Pop~III starburst in this scenario is uncertain\footnote{This co-evolution might provide potential sites for the earliest phase of the Little Red Dots (LRDs).}.
\par
In addition to strong LW radiation feedback, survival of the metal-free gas pocket nearby massive sources is required.
Although we have examined the most massive neighbor within the high-resolution zoom-in region, we did not assign the strength of the SN feedback and metal enrichment in a self-consistent way, to correspond with the high LW-flux levels considered here.
Therefore, we cautiously note that our simulations cannot fully reproduce the favored condition for the Pop~III starburst. However, previous theoretical works offer valuable insights to address the limitations regarding external metal enrichment. For instance, \citet{Agarwal2012} investigated DCBH formation sites at $z>6$, which require a similar strength of LW flux to our Pop~III starburst scenarios. They conclude that pristine halos that are exposed to high LW radiation in the broad mass range considered here can persist until $z > 6$.
Moreover, \citet{Agarwal2019} suggested that high LW radiation ($J_{\rm LW} \geq 10^3 J_{21}$) from nearby massive galaxies can reach distances beyond $10\,\rm kpc$, allowing for a time window during which a Pop~III starburst can occur prior to external metal enrichment.
\par
While there is common agreement on a top-heavy IMF for Pop~III stars compared to Pop~II and Pop~I, the exact shape and mass range of the primordial IMF remain uncertain (\citealp{Klessen2023}). Furthermore, \citet{Sharda2025} suggested that magnetic fields can act to limit the mass that can be reached by individual stars within a Pop~III star cluster to $M_{\rm max} \lesssim 65 \msun$. Adopting different assumptions for the Pop~III IMF will affect our predictions for the transition timescale from Pop~III to Pop~II star formation, as well as the observability of Pop~III stars during the starburst phase.
Additionally, \citet{Storck2025} found that some of the halos in their simulations were able to preserve pristine gas clouds even after the formation of the first Pop~III stars, leading to large final Pop~III masses: this occurred when stars formed within the progenitor mass range for directly collapsing into BHs at the end of their life, with no significant contribution to metal enrichment \citep[e.g.,][]{Heger2003}. Finally, we have also verified that different assumptions on the detailed Pop~III IMF do not significantly impact our post-processing detectability analysis.

\subsection{Effect of cosmic reionization}
\label{subsection:Reionization}
\begin{figure}
    \centering
    \includegraphics[width=85mm]{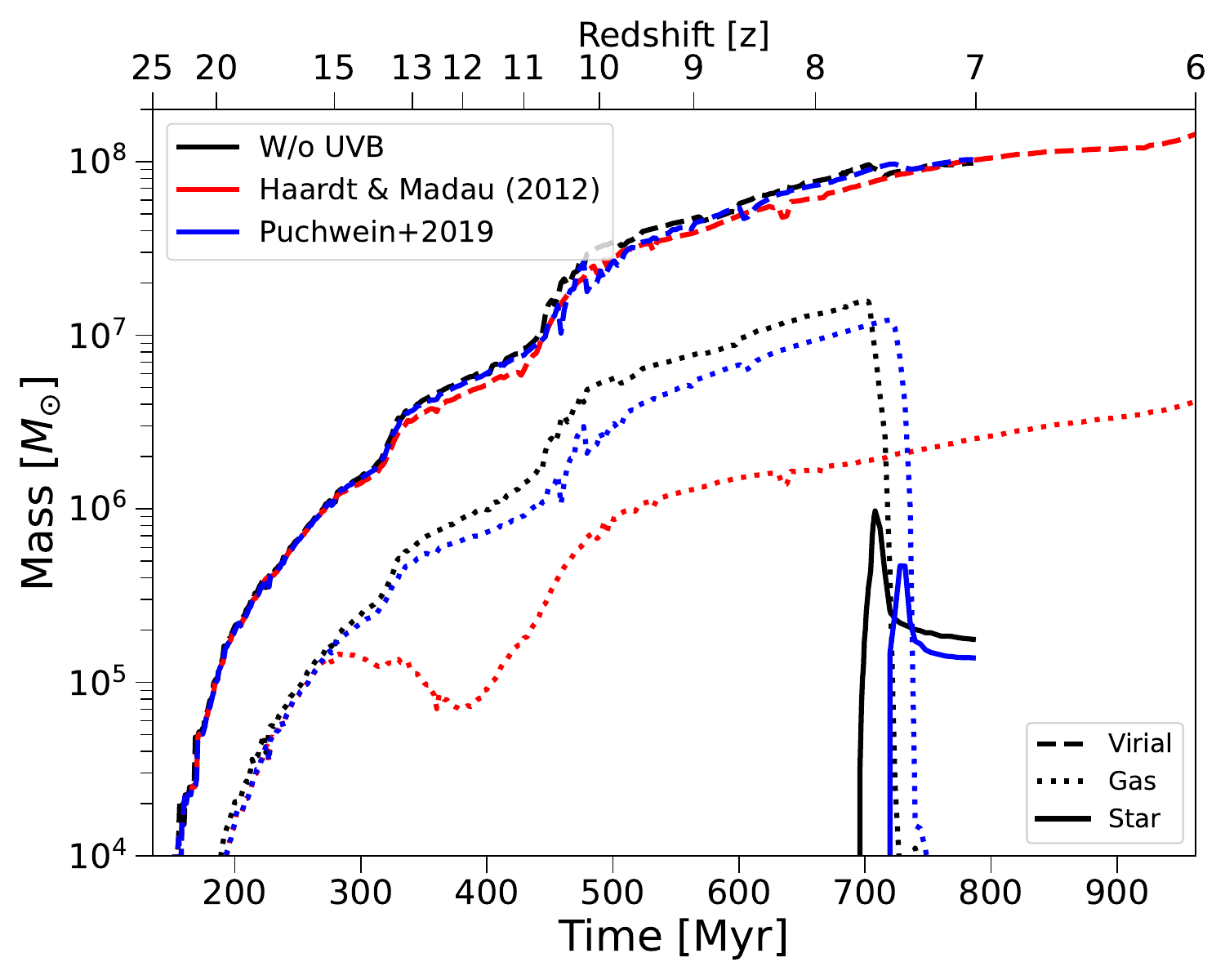}
    \caption{Mass evolution within our simulated galaxy for the LW1e3E100 set as a function of redshift, without (black) and with an external UV background, modeling the effect of reionization, according to  \citet{Haardt2012} (red) and \citet{Puchwein2019} (blue). We adopt the same convention for the lines as in Fig.~\ref{fig:mass_evolution}. For the early reionization scenario resulting from the \citet{Haardt2012} UV background, our simulated galaxy fails to form a Pop~III starburst and suppresses star formation until the simulation ends ($z = 6$). However, for the rapid reionization scenario of \citet{Puchwein2019}, a Pop~III starburst has been triggered with only a slight delay of $\Delta t \approx 25 \rm \,Myr$, compared to the no-UV background case.}
    \label{fig:reionization}
\end{figure}
Owing to the patchy nature of reionization, the onset and progression of reionization can vary significantly across different environments \citep[e.g.,][]{Kim2023, Zier2025b}. Even in the presence of strong LW radiation from nearby sources, ionizing photons from the global UV background can significantly influence the star formation history of our simulated galaxies, in particular by delaying the onset of Pop~III starbursts \citep[e.g.,][]{Borrow2023}. Therefore, taking the LW1e3E100 run without ionizing UV background as our fiducial model, we additionally explore two reionization scenarios by adopting the early reionization UV background model of \citet{Haardt2012} and the rapid-evolution model of \citet{Puchwein2019} \footnote{For the detailed comparison, we refer readers to figure~6 of \citet{Puchwein2019}.}.
\par
In Figure~\ref{fig:reionization}, we present the resulting mass evolution of the simulated galaxies. In the early reionization scenario, $\rm H_{2}$ formation inside the simulated galaxy is already strongly suppressed by the intense LW background, allowing ionizing photons from the UV background to efficiently photo-heat gas clouds while the halo still remains within the molecular-cooling (minihalo) mass regime. As a result, gas clouds are effectively expelled from the halo, and subsequent gas accretion is suppressed through Jeans filtering \citep[e.g.,][]{Quinn1996,Thoul1996,Okamoto2008,Finlator2017}. Consequently, star formation in the simulated galaxy is suppressed until $z = 6$, when the simulation is terminated.
\par
In contrast, the rapid reionization model produces evolution broadly similar to that of the fiducial run without a hard UV background. Although ionizing radiation from the UV background delays the onset of the starburst by $\Delta t \approx 25 \,\rm Myr$, the overall evolutionary history remains comparable to the fiducial case. These results suggest that ionizing photons associated with cosmic reionization can substantially influence the formation of Pop~III starburst galaxies at $z\lesssim 7$, leading to qualitatively different outcomes depending on the adopted reionization history. We therefore cautiously note that the timing and evolution of reionization critically affect our results, effectively determining whether the system experiences complete suppression of star formation or undergoes a Pop~III starburst episode by $z \sim 6$.
\par
Nonetheless, our fiducial simulations without hard UV background provide general insight into the formation of Pop~III starburst galaxies, particularly in the high-redshift Universe ($z \gtrsim 10$). For example, \citet{Maiolino2026} and \citet{Ubler2026} reported the detection of strong HeII $1640$\AA\, emission from the {\it Hebe} object, located near GN-z11. Furthermore, \citet{Ubler2026} confirmed the presence of the H$\gamma$ line without detecting accompanying metal lines, making {\it Hebe} one of the most promising Pop~III galaxy candidates identified to date. Still, there is a key open question about how pristine gas could survive in the vicinity of GN-z11, whereas {\it Hebe}'s environment broadly resembles the conditions inferred from our simulations in terms of exposure to an intense LW radiation field generated by nearby massive sources \citep{Jeon_Hebe2026}.

\section{Summary and Conclusions}
\label{sum&Con}

Motivated by recent JWST observations, we investigated the conditions that may produce large Pop~III starbursts in atomic-cooling halos at the end of the EoR, as well as their detectability.
To find the highest mass range that can be reached by these starbursts, we conducted cosmological radiation-hydrodynamic zoom-in simulations, targeting halos in the lower end of the atomic-cooling halo mass range ($M_{\rm vir} \approx 10^8 \msun$ at $z \approx 7$). We varied the strength of the LW background up to $\bar{J}_{21,0} = 10^4$ -- mimicking the integrated LW radiation field from multiple local sources -- in order to explore the expected delay in the onset of star formation. We further varied the assumed star-formation efficiency at the cloud scale, with the highest value ($\epsilon_{\rm ff} = 1.0$) corresponding to a scenario in which star formation occurs in a short burst ($\Delta t_{\rm burst} < 5 \rm Myr$). Our findings suggest that galaxies exposed to a high LW flux ($J_{\rm LW} \geq 10^3J_{21}$) experience a strong burst of Pop~III star formation ($M_{\star, \rm Pop~III} > 10^5 \msun$), which would be detectable within JWST surveys enforcing gravitational lensing.
\par
We summarize our main results below: 
\begin{itemize}
    \item[$\bullet$] High LW backgrounds ($\bar{J}_{21,0} \geq 10^3$) can delay the first star-formation episode within the simulated galaxies, resulting in the formation of Pop~III clusters with a total mass in the range $10^5 \msun\leq M_{\rm \star, Pop~III} \leq10^6\msun$.
    \item[$\bullet$] For $\bar{J}_{21,0} \geq 10^3$, we find a bifurcation into hot ($T_{\rm gas} \gtrsim10^4 \rm \,K$) and cold ($T_{\rm gas} < 10^4 \rm \,K$) phases for the dense ($n_{\rm H} \gtrsim10 \rm cm^{-3}$) gas located at the center of the galaxies. The central gas exhibits a Matryoshka-like structure, with the cold dense region -- where $\rm H_{2}$ self-shielding is activated -- enclosed in a hot-gas layer heated by LW feedback, and Pop~III starbursts can occur within the cold dense region.
    \item[$\bullet$] We tracked the metal-ejecta from nearby star-forming halos that might externally enrich our target galaxy. We find that the Pop~III starburst phase in galaxies with a high LW background is initiated before the metal bubble reaches the halo, so that the first metals come from internal enrichment instead. This confirms that our simulated starburst galaxies are found within pristine gas pockets, in agreement with a patchy scenario for global metal enrichment. 
    \item[$\bullet$] By modeling their emission, we find that galaxies with $\bar{J}_{21, 0} = 10^3\,(10^4)$ can reach magnitudes of $m_{\rm AB} \lesssim 32 \, (31)\rm \, mag$ during the initial Pop~III starburst phase, which are detectable in current JWST surveys leveraging gravitational lensing (e.g., GLIMPSE) throughout the entire starburst phase, and detectable in non-lensed surveys during initial Pop~III burst for $\bar{J}_{21, 0} = 10^4$.
\end{itemize}

Our findings suggest that galaxies immersed in a high LW flux from local sources present ideal conditions for the formation of the most intense Pop~III starbursts with $M_{\rm \star, Pop~III} \lesssim 10^6\msun$.
The conditions favoring intense Pop~III starbursts at the end of the EoR are expected to be rare, due to the stringent requirements of both a high LW intensity from nearby, massive systems, and the persistence of metal-free pockets. 
We plan to extend our study to larger simulation boxes, capturing both the potential sites for the formation of Pop~III starbursts and nearby, massive systems responsible for producing the LW flux.
\par
We emphasize that detecting Pop~III galaxies will be challenging, but not impossible. Particularly, targeting massive halos in overdense regions of the cosmic web during the EoR appears a promising strategy, allowing us to select favorable (albeit rare) environments for efficient Pop~III detection. JWST surveys covering large volumes and supported by gravitational lensing, such as VENUS, and surveys with strong lensing fields, such as  Abell 2744 (e.g., UNCOVER, \citealt{Weaver2024}; GLASS, \citealt{Treu2022}; ALT, \citealt{Naidu2024}; MEGASCIENCE, \citealt{Suess2024}), can also provide additional opportunities to detect Pop~III galaxies and unveil previously hidden aspects of early Universe evolution.

\section*{acknowledgements}
We are grateful to Volker Springel, Joop Schaye, and Claudio Dalla Vecchia for permission to use their versions of \textsc{gadget}.
Also, we thank Junehyoung Jeon and Juyoung Kim for helpful comments.
The authors acknowledge the Texas Advanced Computing Center (TACC) for providing HPC resources under allocation AST23026. A.~V. acknowledges funding from the Cosmic Frontier Center and the University of Texas at Austin’s College of Natural Sciences. M.~J. is supported by Samsung Science and Technology Foundation under Project Number SSTF-BA2402-03.

\bibliographystyle{aasjournal}

\appendix


For comparison with the LW1e3E100 run, we show the gas phase-diagram for LW10E100.
Before the onset of star formation, the high-density gas ($n_{\rm H, thr} \geq 10 \rm \, cm^{-3} $) evolves monotonically, showing only the cooling phase associated with the $\rm H_{2}$ self-shielding effect. We also show the projected temperature, exhibiting a significant distinction between the hot and cold phases.

\begin{figure*}[h]
    \centering
    \includegraphics[width=130mm]{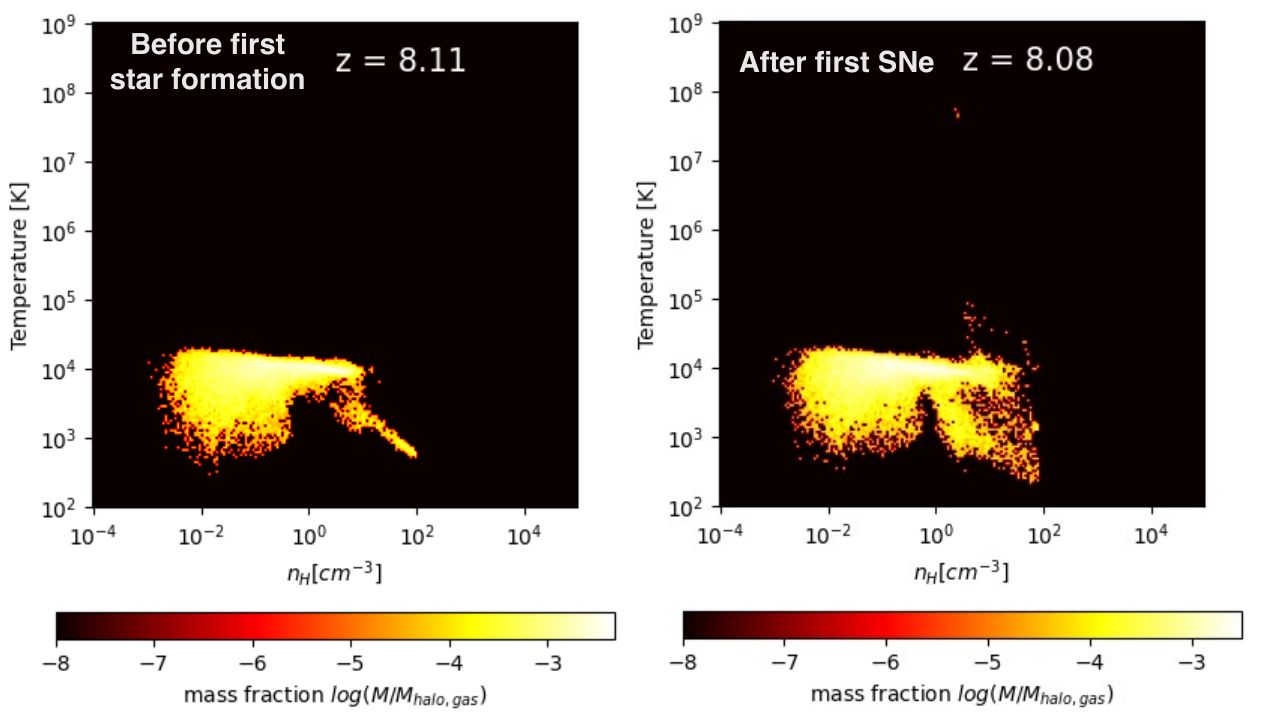}
    \caption{Gas properties within the LW10E100 target galaxy. The left panel exhibits properties of gas particles before the onset of the first star formation, while the right panel shows properties after the first SNe have occurred (as in Fig.~\ref{fig:Phase-diagram}). With lower LW flux, the hot branch of the diagram is suppressed, and the gas above $n_{\rm H} \sim 10 \cmci$ quickly evolves towards lower temperatures.}
    \label{fig:Gas-phase-APX}
\end{figure*}

\begin{figure*}[h]
    \centering
    \includegraphics[width=80mm]{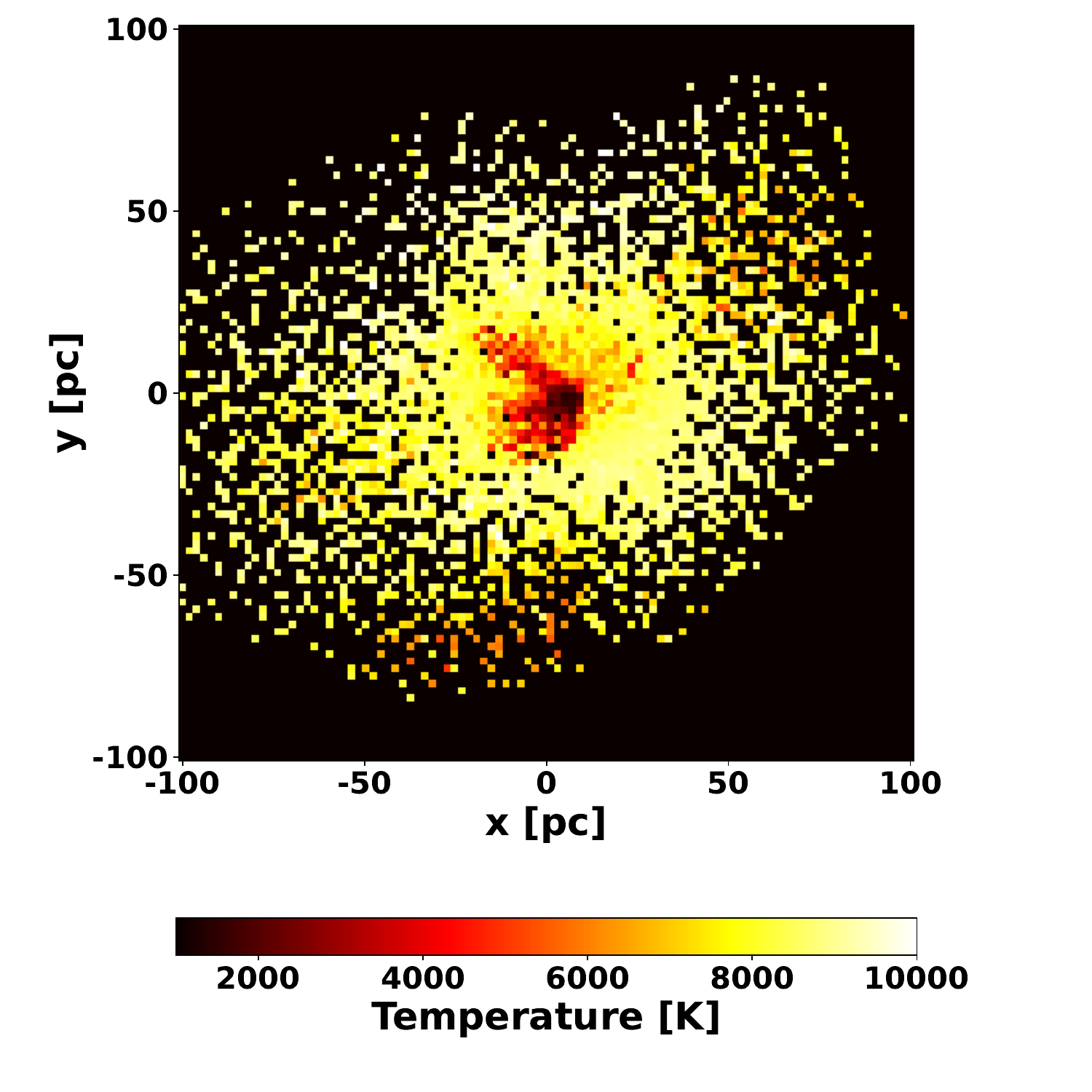}
    \caption{Temperature slice within $0.1R_{\rm vir}$ for the LW1e3E100 run, when the bifurcation inside the phase-tracks begins to emerge ($z\approx7.69$, see Fig.~\ref{fig:Phase-diagram}). 
    We here only consider high-density ($n_{\rm H} \geq 10 \cmci$) gas within the core, revealing a significant temperature drop towards the halo center.
    }
    \label{fig:sliceplot}
\end{figure*}

\begin{table}[h]
    \renewcommand{\arraystretch}{1.3}
    \centering
    \begin{tabular}{c|c|c|c|c|c}
         \hline
         \textbf{Set name}
         & \parbox[c]{2.5cm}{\centering \textbf{$\rm \log L_{H\alpha}$} \\ {\scriptsize [erg s$^{-1}$]}}
         & \parbox[c]{2.5cm}{\centering \textbf{$\rm \log L_{H\beta}$} \\ {\scriptsize [erg s$^{-1}$]}}
         & \parbox[c]{2.5cm}{\centering \textbf{$\rm \log L_{H\gamma}$} \\ {\scriptsize [erg s$^{-1}$]}}
         & \parbox[c]{2.5cm}{\centering $\rm \log L_{HeII \, 1640}$  \\ {\scriptsize [erg s$^{-1}$]}}
         & \parbox[c]{2.5cm}{\centering $\rm \log L_{[OIII]}$ \\ {\scriptsize [erg s$^{-1}$]}} \cr
         \hline \hline
         {\sc LW1e3E100} & 41.20 / 38.81 / 39.43 & 40.53 / 38.35 / 38.95 & 40.14 / 38.03 / 38.63 & 40.50 / 39.18 / 38.00 & - / 35.45 / 38.08 \cr
         \hline
         {\sc LW1e4E100} & 41.58 / 41.25 / 40.22 & 40.94 / 40.63 / 39.68 & 40.56 / 40.24 / 39.34 & 40.85 / 40.68 / 38.60 & - / - / 38.19 \cr
         \hline
    \end{tabular}
    \caption{The luminosity of emission lines of Pop~III starburst galaxies in Fig.~\ref{fig:observation}. Each row shows a different simulation set, LW1e3E100 and LW1e4E100, respectively, while each column shows different emission lines $\rm H\alpha$, $\rm H\beta$, $\rm H\gamma$, $\rm HeII$, and $\rm [OIII]$ doublet. Each cell shows flux density values in three phases: initial Pop~III / mid-starburst / second starburst phase with Pop~II stars, represented by black / blue / red in Fig.~\ref{fig:observation}. }
    \label{tab:line_luminosity}
\end{table}

\end{document}